\documentclass[aps,onecolumn,showpacs,nofootinbib]{revtex4}
\usepackage{graphicx,amsmath,slashed,bbold,subfigure}
\usepackage[latin1]{inputenc}

\def\cc{\c{c}}

\def\ii{\'{\i}}

\newcommand{\ud}{\mathrm{d}}

\begin{document}

\title{Thermodynamic potential with correct asymptotics for PNJL model.}

\author{ J. Moreira, B. Hiller, A. A. Osipov{\footnote{On leave from 
         Dzhelepov Laboratory of Nuclear Problems, 
         Joint Institute for Nuclear Research, 
         141980 Dubna, Moscow Region, Russia}}, 
         A. H. Blin}
\affiliation{Centro de F\'{\i}sica Computacional, Departamento de
         F\'{\i}sica da Universidade de Coimbra, 3004-516 Coimbra, 
         Portugal}



\begin{abstract}
An attempt is made to resolve certain incongruities within the Nambu -- Jona-Lasinio (NJL) and Polyakov loop extended NJL models (PNJL) which currently are used to extract the thermodynamic characteristics of the quark-gluon system.
It is  argued that the most attractive resolution of these incongruities is the possibility to obtain the thermodynamic potential directly from the corresponding extremum conditions (gap equations) by integrating them, an integration constant being fixed in accordance with the Stefan-Boltzmann law.
The advantage of the approach is that the regulator is kept finite both in divergent and finite valued integrals at finite temperature and chemical potential. The Pauli-Villars regularization is used, although a standard 3D sharp cutoff can be applied as well.
\end{abstract}


\maketitle


\section {Introduction}
\vspace{0.5cm}

One of the most successful and widely used approaches to model the QCD phase diagram in the chiral restoration regime is the NJL Lagrangian \cite{Nambu:1961} with the quark condensate as order parameter in the chiral limit, and its extension to incorporate the Polyakov loop, the PNJL model \cite{Fukushima:2004}-\cite{Gatto:2010}, where the traced Polyakov loop is an order parameter in the quenched limit and an indicator of deconfinement.  For realistic mass values of the light and strange quarks  the order parameters are approximate quantities, since the chiral symmetry and center symmetry are explicitly broken. The estimates for the transition temperatures from the hadron to the quark gluon phase range from $T_c\sim 150 - 200$ MeV, according to recent lattice calculations, see e.g. \cite{Cheng:2006}-\cite{ DeTar:2009}. The nature of the transition as function of the temperature and baryonic density, the eventual coincidence of chiral and deconfinement temperatures, the role played by the strange quark, the possible existence of a critical endpoint (CEP) and the symmetry breaking pattern associated with its location, are some of the topics which are currently under intense study at experimental facilities (GSI, RHIC, CERN, LHC), in lattice calculations and effective model approaches, see e.g. \cite{Kogut:2004} - \cite{Fukushima:2010} for recent reviews.   

The NJL model was originally introduced as an interesting field theoretical illustration of the dynamical chiral symmetry breaking phenomena in particle physics. With years it became clear that the effective quark interactions capture reasonably well the main features of the low energy meson dynamics. The thermodynamics of the model is nowadays extensively used with the hope that some of the basic features of the hot and dense QCD matter are well described (at least in the chiral symmetry restoration regime), and this hope has definite grounds (see reviews \cite{Shuryak:2004}, \cite{Buballa:2005}).
At the same time, the NJL model thermodynamics, as it now stands, has certain internal problems, related with the cutoff dependence of the results. This is a general drawback feature of any non renormalizable field theory model. As a consequence, the asymptotic behavior of some relevant thermodynamic quantities is in conflict with some of the well known general laws. The thermodynamic potential separates naturally into the vacuum piece and the temperature dependent matter part. It has been reported that the asymptotic description of several quantities associated with the thermodynamic potential cannot be achieved using the same regularization criteria on vacuum and matter integrals for all observables. For instance, the effective number of quark degrees of freedom has the correct asymptotics at large $T$ (Stefan-Boltzmann limit) only when the cutoff is removed from the finite matter contributions \cite{Klevansky:1994}. On the other hand the removal of the regulator from the matter pieces leads asymptotically to unphysical positive condensates of the quarks, except in the chiral limit. In other words the  explicit symmetry breaking pattern will be restored asymptotically only if the regulator is also kept in the matter integrals.
 
In the present contribution we argue that it is possible to achieve consistent regularization criteria for vacuum and matter integrals for these observables simultaneously, by obtaining the thermodynamic potential through integration of the model gap equations \cite{Hiller:2010},\cite{Coleman:1973}. The integration runs starting from zero over the whole spectrum of the auxiliary scalar fields $M$ and is defined up to an integration constant which can be  $T,\mu$ dependent. We fix this constant in a regulator independent way, by requiring that the low 
$T,\mu$ behavior of the thermodynamic potential yields the correct number of degrees of freedom.

Then we extend this method of obtaining the thermodynamic potential to the PNJL model. With the present treatment we obtain the correct model asymtotics for all observables considered: the number of degrees of freedom, the traced Polyakov loop and the dynamical quark masses (or condensates).
This is in contrast to the standard approach which requires, as in the case of the NJL model, the removal of the regulator from the finite integrals to describe the number of degrees of freedom, and its presence for the remaining observables; for example the cutoff must be kept to describe the correct asymptotics of the traced Polyakov loop \cite{Sazaki:2007}. Although for certain Polyakov potentials, of logarithmic form \cite{Fukushima:2008},\cite{Weise:2006}, it is still possible to display its correct asymptotics without a regulator in the matter integrals, the physical values of the condensates at high $T$ necessarily require its presence.

The thermodynamic properties of the NJL and PNJL model have most widely been analyzed regularizing it with a 3D sharp cutoff. This has prompted us to use a different regularization to try to clarify  the impact that the choice of a regulator may have on the asymptotics. We take the Pauli-Villars regularization (PV) \cite{PV:1949} which allows to check for effects associated with covariance and can be readily used in extensions of the model involving the vector mesons, where it is recommendable to use. We take the PV regularization kernel of \cite{Osipov:1985}, which to our knowledge was used for the first time in the NJL model. It has the property that the scalar quadratic and logarithmic divergent amplitudes coincide with the ones of the covariant sharp cutoff.  Since all the PV regulated integrals can be written in 3-momentum space with 3-momentum dependent regulators, it is easy to trace the differences between the two methods. This comparison between regularizations will lead us to the conclusion that if one follows the steps indicated above to derive the thermodynamic potential, both the PV and $3D$ sharp cutoff regularizations can be used to achieve the correct asymptotic behavior.

The paper set up is as follows. After a brief review in section II of the PV regularized fermion one-loop determinant for the NJL thermodynamic potential obtained in \cite{Hiller:2010}, which we use to introduce the notation,
we proceed in section III to regularize the PNJL model with the PV regularization. The large $T$ asymptotic behavior of several thermodynamic quantities is discussed in section IV. Numerical results for the effective number of degrees  of freedom, gap equation solutions for the constituent quark masses and Polyakov loop fields are given in Section V, using several parametrizations for the Polyakov potential  and parameter sets for the many-quark Lagrangian. These results are compared with the conventional way of calculating with the 3D cutoff. Conclusions are presented in Section VI. In the Appendix C we discuss how the present procedure can be adapted to the 3D sharp cut-off case, leading to compatible results for both types of regularization applied on the amplitudes considered.

\section{PV regularization of the NJL model}

In a recent investigation \cite{Hiller:2010} we have proven and discussed in great detail how to obtain the correct asymptotics of the effective number of quark degrees of freedom at large $T$ for the NJL model, using the PV regularization applied to vacuum and matter integrals. This study was done for the NJL model extended to include the 't Hooft $2 N_f$ ( $N_f=$ number of flavors) determinant interaction  \cite{Hooft:1976},\cite{Bernard:1988},\cite{Reinhardt:1988} related with the axial $U(1)_A$ anomaly, and the eight quark interaction Lagrangian \cite{Osipov:2005b}. The latter is proven necessary to render the ground state of the theory stable in the case of $N_f=3$ \cite{Osipov:2006}, \cite{Osipov:2006a},\cite{Osipov:2001}. In the text we refer to these extensions of the NJL model as NJLH and NJLH8 respectively (whereas the Polyakov extensions are referred to as PNJLH and PNJLH8) \footnote{In some occasions we will use PNJL to refer collectively to these extensions but we hope it will be clear from the context.}. 
The resulting thermodynamic potential as function of the three independent variables $M_f=\{M_u,M_d,M_s\}$ and $T,\mu$ was derived to be \cite{Hiller:2010}
\begin{eqnarray} 
\label{effpot-t}
     V(M_f,T,\mu )
     &\!\! = \!\!&\frac{1}{16}\left.\left(4Gh_f^2+\kappa h_uh_dh_s 
     +\frac{3g_1}{2}\left(h_f^2\right)^2+3g_2h_f^4\right)\right|_0^{M_f} \nonumber \\
     &\!\! +&\frac{N_c}{8\pi^2}\!\sum_{f=u,d,s}\!\! J_{-1}(M_f^2,T,\mu ) 
     + C(T,\mu ),
\end{eqnarray}
through integration of the gap-equations:  
\begin{equation}
\label{gap-t}
            h_f + \frac{N_c}{2\pi^2} M_f J_0(M_f^2,T,\mu )=0.
\end{equation}
These are solved self-consistently with the stationary phase conditions (SPA)
\begin{align}
\label{StaEq}
\left\{
\begin{array}{l}
m_u-M_u=G h_u +\frac{\kappa}{16}h_d h_s +\frac{g_1}{4}h_u h^2_f+\frac{g_2}{2}h_u^3\\
m_d-M_d=G h_d +\frac{\kappa}{16}h_u h_s +\frac{g_1}{4}h_d h^2_f+\frac{g_2}{2}h_d^3\\
m_s-M_s=G h_s +\frac{\kappa}{16}h_u h_d +\frac{g_1}{4}h_s h^2_f+\frac{g_2}{2}h_s^3
\end{array}
\right.
\end{align}
to yield the constituent quark masses $M_f$. Here $m_f$ are the current quark masses, $h_f= 2 <0|{\bar f} f|0>$, ($f=\{u,d,s\}$), is defined as twice the quark condensate in the symmetry broken phase,  and we use the abbreviations $h_f^2=h_u^2+h_d^2+h_s^2, \,\, h_f^4=h_u^4+h_d^4+h_s^4$, and  $G,\kappa,g_1,g_2$ denote respectively the $4q, 6q$ and the two $8q$ coupling strengths, of which $g_1$ corresponds to a OZI-violating combination. Given the following specific constraints among the couplings, $g_1 > 0, g_1+3 g_2 > 0, G>\frac{1}{g1}(\frac{\kappa}{16})^2$, there exists only a single real solution to the cubic SPA equations \cite{Osipov:2005b} 
and the global stability of the effective potential is guaranteed
. For simplicity the case with $\mu=\mu_u=\mu_d=\mu_s$ was considered, generalization to account for nonzero isospin chemical potential can be done as e.g. in \cite{Klimenko:2006}.
The quark one loop integrals in (\ref{gap-t}) are given by:
\begin{eqnarray}
\label{J0t}
    &&J_0(M^2,T,\mu )=4\int_0^\infty\ud |\vec{p}_E| |\vec{p}_E|^2
     \hat{\rho}_{\Lambda\vec{p}_E}\frac{1}{E_p}\left(1-n_{q}-n_{\overline{q}}\right) \nonumber \\
    &&=J_0(M^2)- 4\int_0^\infty\ud |\vec{p}_E| |\vec{p}_E|^2
     \hat{\rho}_{\Lambda\vec{p}_E}\frac{n_{q}+n_{\overline{q}}}{E_p},
\end{eqnarray}
where
\begin{equation}
J_0(M^2)=4\int_0^\infty\ud |\vec{p}_E| |\vec{p}_E|^2
     \hat{\rho}_{\Lambda\vec{p}_E}\frac{1}{E_p}= \Lambda^2-M^2\ln\left(1+\frac{\Lambda^2}{M^2}\right)
\end{equation}
describes the vacuum contribution. 
The integral $J_{-1}(M^2,T,\mu )$ resulting from the integration over $M$, 
\begin{equation}
\label{J-1}
     J_{-1}(M^2,T,\mu )=-\int_0^{M^2}\!\!\! J_0(M^2,T,\mu )\ud M^2
     =J_{-1}(M^2)+J^{\mathrm{med}}_{-1}(M^2,T,\mu ),
\end{equation}
splits into a vacuum,
\begin{eqnarray}
\label{J_1}
     &&J_{-1}(M^2)=-\int_0^{M^2}\!\! J_0(M^2)\ud M^2 = -8 \int_0^{\infty}\ud |\vec{p}_E| |\vec{p}_E|^2 \hat{\rho}_{\Lambda\vec{p}_E} (E_p(M) -E_p(0)) \nonumber \\
     &&=\frac{8}{3} \int^\infty_0\ud 
     |\vec{p}_E||\vec{p}_E|^4\hat{\rho}_{\Lambda\vec{p}_E}\left(
     \frac{1}{E_p(M)}-\frac{1}{E_p(0)}\right)
     =-\frac{1}{2}\left(M^2J_0(M^2)+\Lambda^4\ln
     \left(1+\frac{M^2}{\Lambda^2}\right)\right),
\end{eqnarray}
and a medium contribution,
\begin{eqnarray}
\label{J-1med}
     J_{-1}^{\mathrm{med}}(M^2,T,\mu )
     &\!\! =\!\!&8T\int^\infty_0\!\ud |\vec{p}_E||\vec{p}_E|^2
     \hat{\rho}_{\Lambda\vec{p}_E}\ln\frac{\left(1+e^{-\frac{E_p(0)-\mu}{T}}\right)
     \left(1+e^{-\frac{E_p(0)+\mu}{T}}\right)}{\left(1+e^{-\frac{E_p(M)-\mu}{T}}
     \right)\left(1+e^{-\frac{E_p(M)+\mu}{T}}\right)}
     \nonumber \\
     &\!\! =\!\!& -\frac{8}{3}\int^\infty_0\ud 
     |\vec{p}_E||\vec{p}_E|^4\hat{\rho}_{\Lambda\vec{p}_E}\left(
     \frac{n_{qM}+n_{\overline{q}M}}{E_p(M)}-
     \frac{n_{q0}+n_{\overline{q}0}}{E_p(0)}\right).
\end{eqnarray}
To obtain the second equality a partial integration has been performed, the boundary term vanishes for the PV regularization.
The notation $n_{q0}, n_{\overline{q}0}$, and $n_{qM}, n_{\overline{q}M}$ refers to the quark and antiquark
occupation numbers for massless and massive particles correspondingly, $E_p(M)
=\sqrt{M^2+\vec{p}^{\, 2}_E}$, $E_p(0)=|\vec{p}_E|$, generically defined as:
\begin{equation}
\label{n-occup}
     n_{q}=\frac{1}{1+e^{\frac{E_p-\mu}{T}}}, \quad 
     n_{\overline{q}}=\frac{1}{1+e^{\frac{E_p+\mu}{T}}}.
\end{equation}
The Pauli-Villars regulazion kernel that we employ,
\begin{equation}
\label{PV}
     \hat{\rho}_{\Lambda\vec{p}_E}
     =1-\left(1-\Lambda^2 \frac{\partial}{\partial \vec{p}_E^{\, 2}}\right)
     \exp\left(\Lambda^2\frac{\partial}{\partial\vec{p}_E^{\, 2}}\right),
\end{equation}
corresponds to two subtractions in the integrand \cite{Osipov:1985}. 

The PV result has some unconventional features. First note the occurrence of occupation numbers for massless particles in (\ref{J-1med}). These are due to the thermodynamic potential being a direct extension to finite $T,\mu$ of the effective potential which is normalized to $V(M_f=0,0,0)=0$  and  implies the zero mass energy subtraction in the vacuum piece (\ref{J_1}), \cite{Osipov:2004a}, \cite{Hiller:2010}. As opposed to a 3D sharp cutoff, this subtraction in (\ref{J_1}), which is a natural consequence of requiring the power series expansion of the effective potential to start off at zero when the quark masses vanish, is also necessary in the PV regularization scheme to render the 3-momentum integration finite, despite the presence of the regulator. Once one accepts the evidence that thermodynamic consistency cannot be achieved if only the vacuum is regularized (shown in the following sections), the massless occupation numbers must be fully considered in the regularization of the matter parts. 

Second there occurs the term  $C(T,\mu)$, with the following significance. The degrees of freedom associated with the massless occupation numbers do not vanish in the $T\rightarrow 0$ limit. We use a freedom associated with an integration constant associated to the integration of the mass gap equations to impose the boundary condition that as $T\rightarrow 0$ the correct limit is obtained. This leads to the $M$ independent quantity 
\begin{equation}
\label{FTP}
       C(T,\mu )= -\frac{N_c}{\pi^2}\int_0^\infty\ud |\vec{p}_E||\vec{p}_E|^4
       \frac{n_{q0}+n_{\bar q0}}{E_p(0)}\ \to\ C(T,0)
       =-\frac{7N_cN_f}{180}\pi^2T^4.
\end{equation}
Since at low $T$ the matter integrals depending on non-vanishing quark masses (and thus also on the PV regulating masses) vanish exponentially \cite{Hiller:2010}, $C(T,\mu)$ is regularization independent and coincides with the corresponding regulated expression ($\hat{\rho}_{\Lambda\vec{p}_E}\to 1$). The matter quark-loop part of the thermodynamic potential at finite $T,\mu$, i.e. the $J^{\mathrm{med}}_{-1}$-part of the thermodynamic potential, together with the boundary condition term $C(T,\mu)$, coincides in the low $T$ regime with the matter part obtained with the standard approach when the 3D sharp cutoff is removed  ($\Lambda_3\to\infty$). 

It is remarkable that the large $T$ asymptotics is as well correctly described. The reason is that at high $T$ the regularization of the matter parts becomes relevant. This is taken into account in the PV regulated thermodynamic potential (\ref{effpot-t}). In other words we have shown that combining the $C(T,\mu)$ term with the regularized expression for massless contributions eliminates completely any contribution of massless states to the number of degrees of freedom. What remains are the unphysical states associated with the corresponding PV regulating masses which in turn are cancelled by the PV regulators of the massive states at leading order of the high T expansion. Thus all auxiliary fields introduced in the regulating process are absent in the asymptotic description, leading to the model parameter independent and correct result for the number of degrees of freedom. This was proven in all detail in \cite{Hiller:2010}.

In the present work we adopt a different but equivalent way of looking at these cancellations. We separate the term $C(T,\mu)$, which does not depend on the regulator, from the remaining regulator dependent matter pieces. Doing so we show in section 4 and in the Appendix A that the regularized contributions lead in the high $T$ limit to the vacuum expression, up to a sign. The full cancellation of the regularized matter and vacuum pieces at high $T$ allows to understand easily the asymptotic behavior of the thermodynamic potential and gap equations. Superficially seen this way to group terms may create the illusion that the $C(T,\mu)$ is an unphysical Stefan-Boltzman contribution associated with massless terms. It is not, since the regularized expression itself contains massless terms which cancel it. This is very easily understood by considering the expression (\ref{equ}) of the Appendix. The leading terms in the high $T$ expansion for the massless and massive terms are equal to each other and to the $C(T,\mu)$ term. Their roles can be trivially interchanged. This result is the simple consequence that at high $T$ terms of the order $M^2/T^2$ can be neglected.  

We recall that in the case with 3D cutoff the $M=0$ subtractions are not performed and the regulator is a step function $({\hat \rho}_{\Lambda_3}= \Theta(\Lambda_3-|\vec p_E|))$. Besides this the integrand of our medium contribution  $J_{-1}^{\mathrm{med}}(M^2,T,\mu )$ differs by a partial integration from the $3D$ cutoff one, since the corresponding surface term vanishes in the PV case, as opposed to the $3D$ case. 

\section {PV regularization for the PNJL model}
 
We are now ready to extend our analysis to include the Polyakov loop dynamics \cite{Fukushima:2004}-\cite{Gatto:2010},\cite{Fukushima:2008},\cite{Weise:2006}, (in the present work we do not consider the dressed Polyakov loop case, see e.g \cite{Kashiwa:2009}- \cite{Gatto:2010}). This is effected 
by coupling the quark and gluonic fields in the temporal gauge through the covariant derivative  $\partial^\mu\rightarrow D^\mu=\partial^\mu+\imath A^\mu$, with $A^\mu=\delta^\mu_0 g A^0_{a}\frac{\lambda^a}{2}$ ($\lambda_a$, $a=1,\ldots,8$ are the standard Gell-Mann matrices in color space) and by assuming the temporal component in euclidean space-time  $A_4=\imath A^0$ to be static and diagonal $A_4=A_3\lambda_3 + A_8\lambda_8$ (the Polyakov gauge). In addition, a temperature dependent pure gluonic term is introduced through the Polyakov potential $\mathcal{U}\left(\overline{\phi},\phi,T\right)$,  where $\phi$ is the traced Polyakov loop and $\overline{\phi }$ its charge conjugate
\begin{align}
\phi=\frac{1}{N_c}\mathrm{Tr_c} L,\quad \overline{\phi}=\frac{1}{N_c}\mathrm{Tr_c} L^\dag,\quad
L=\mathcal{P} e^{\int^\beta_0 \mathrm{d}x_4 \imath A_4 },
\end{align}
with $N_c$ representing the number of colors and $\mathcal{P}$ denoting path-ordering. In the framework of the model $\phi$ and $\overline{\phi}$ are treated as classical field variables. The form of the potential $\mathcal{U}\left(\overline{\phi},\phi,T\right)$ can be determined, for instance, from lattice calculations in the pure gauge sector. Several parametrizations have been considered and we shall address them below (see Table \ref{PolyakovPot}).

In the Polyakov gauge it has been shown that the phase of the Polyakov loop appears in the quark action in the form of an imaginary chemical potential \cite{Weiss:1981}. This observation allows for a simple extension of the thermodynamic potential through the following steps: 
\vspace{0.5cm}

(i) Replacement of the occupation numbers in (\ref{J-1med}) by the tilde quantities 
\begin{align}
\label{occpol}
\tilde{n}_{qM}\left(E_p,\mu,T,\phi,\overline{\phi}\right)\equiv
&\frac{1}{N_c}\sum_{i}^{N_c}n_{qM} (E_p,\mu+\imath \left(A_4\right)_{ii},T)\nonumber\\
=&\frac{
\left(\phi+2\overline{\phi} e^{-\left(E_p-\mu\right)/T}\right) e^{-\left(E_p-\mu\right)/T}+e^{-3\left(E_p-\mu\right)/T}}
{1+3\left(\phi+\overline{\phi} e^{-\left(E_p-\mu\right)/T}\right)e^{-\frac{E_p-\mu}{T}}+e^{-3\frac{E_p-\mu}{T}}}
.
\end{align}
The equivalent expression for $\tilde{n}_{\overline{q}M}\left(E_p,\mu,T,\phi,\overline{\phi}\right)$ can be obtained by replacing $\mu\leftrightarrow-\mu$ and $\phi\leftrightarrow\overline{\phi}$. In absence of the gauge field, $A_4=0$, one obtains $\phi=\overline{\phi}=1$,  and one recovers the occupation numbers (\ref{n-occup}).
Note that in the presence of  a non-vanishing gauge field the limiting value unity for $\phi,\overline{\phi}$ is also reached as $T\rightarrow \infty$, but then obviously the limit must be taken simultaneously in all terms of $\tilde{n}_{q},\tilde{n}_{\overline{q}}$. 

(ii) Specification of the values of $\phi,\overline{\phi}$ at the integration boundary  $M=0$. On first sight, it might seem natural to systematically implement the substitution (\ref{occpol}) also for the massless occupation numbers. However this would lead to a dependence on $\phi,\overline{\phi}$ of $C(T,\mu )$ which must be field independent in order not to alter the gap equations. Therefore, recalling that the boundary $M=0$ has been introduced at
$T=\mu=0$, in absence of the Polyakov loop, the values of $\phi,\overline{\phi}$ are bound to be unity. 

As a result of these conditions we obtain the final expression for the thermodynamic potential including the Polyakov loop as
\begin{eqnarray} 
\label{effpot-t-polyakov}
     V(M_f,T,\mu,\phi,\overline{\phi} )
     &\!\! = \!\!&\frac{1}{16}\left.\left(4Gh_f^2+\kappa h_uh_dh_s 
     +\frac{3g_1}{2}\left(h_f^2\right)^2+3g_2h_f^4\right)\right|_0^{M_f} \nonumber \\
     &\!\! +&\frac{N_c}{8\pi^2}\!\sum_{f=u,d,s}\!\! J_{-1}(M_f^2,T,\mu,\phi,\overline{\phi} ) + C(T,\mu ) \nonumber \\
     &\!\! +& \mathcal{U}\left(\phi,\overline{\phi},T\right).
     \end{eqnarray}

The minimization of the thermodynamic potential with respect to $M_u$, $M_d$, $M_s$, $\phi$ and $\overline{\phi}$ gives rise to the gap equations.

\section {Asymptotics}

At this stage we are already able to draw the following conclusion about the PV regularization on the asymptotic value of the total number of degrees of freedom $\nu(T)$. 
This quantity at $\mu=0$ is defined as the pressure dif\mbox{}ference from the zero-temperature value, in Stefan-Boltzmann units $\pi^2T^4/90$,
\begin{equation}
\label{nu}
      \nu (T) = \frac{p(T)-p(0)}{\pi^2T^4/90}, 
\end{equation}
with $p(T)=-V(M^*_f,\phi*,\overline{\phi}*,T,\mu =0)$, where the starred values $M^*_f,\phi*,\overline{\phi}*$ denote the gap equation solutions at a given $T$. 
In \cite{Hiller:2010}, where the extension to include the Polyakov loop was not considered, we explain how the asymptotic behavior of this quantity for large $T$ comes completely and correctly determined by $C(T,\mu )$  independently of the model parameters. Now, according to our observation (ii) above,  the covariant derivative coupling of the Polyakov loop to the quarks does not affect $C(T,\mu )$. In the extended version of the model, here considered, once again the asymptotic values for the quark number degrees of freedom at large $T$ is determined by $C(T,\mu )$ whereas the Polyakov potential 
$\mathcal{U}\left(\phi*,\overline{\phi}*,T\right)$ will be the only source of degrees of freedom associated with the Polyakov loop at large $T$; (henceforth we drop the * denoting the gap equation solutions in text and figures, in order to simplify the notation). The reason for this is as follows: in the $T\rightarrow \infty$ limit we obtain that $J_{-1}(M_f^2,T,\mu,\phi,\overline{\phi} )=0$, since then the matter part 
\begin{equation}
J^{med}_{-1}(M_f^2,T\rightarrow \infty,\mu,\phi,\overline{\phi} )= -\frac{8}{3}\int^\infty_0\ud 
     |\vec{p}_E||\vec{p}_E|^4\hat{\rho}_{\Lambda\vec{p}_E}\left(
     \frac{1}{E_p(M)}-
     \frac{1}{E_p(0)}\right),
\end{equation}
is independent of $\phi,\overline{\phi}$ and equal to the negative of the vacuum expression, $J_{-1}(M_f^2)$, eq. (\ref{J_1}). The same happens in the case of the $J_{0}(M_f^2,T,\mu,\phi,\overline{\phi} )$ integrals (\ref{J0t}), which reduce to the vacuum expression in the high $T$ limit. Thus the $h_f$ dependent terms in the pressure also vanish in this limit through the gap equations (\ref{gap-t}).

Therefore we conclude two things. Firstly, since the matter parts become up to a sign identical to their vacuum pieces as $T\rightarrow \infty$,  they need to be regularized to guarantee asymptotic consistency, see also \cite{Florkowski:1997}, \cite{Hiller:2010}. Secondly, the large T asymptotics for the $\phi,\overline{\phi}$ related number of degrees of freedom is fully driven by the potential term $\mathcal{U}\left(\phi,\overline{\phi},T\right)$. This can also be understood reversing the argument, i.e. leaving the medium part unregularized, which consists in discarding the PV subtractions.  In this case, using the following representation \cite{Hiller:2010} (we consider here for simplicity the case with $\phi=\overline{\phi}=1$ and $\mu=0$ since it does not alter the large asymptotic T dependence of the integral) for a term in $J^{med}_{-1}$ (see the Appendix for a more detailed discussion)
\begin{equation} 
\label{int4}
     \int^\infty_0\ud |\vec{p}_E||\vec{p}_E|^4\frac{n_{qM}}{E_p(M)}
     =T^4\int^\infty_0\!\ud x\frac{(x^2+2x\frac{M}{T})^{\frac{3}{2}}}{1+
     e^{x+\frac{M}{T}}}\sim \frac{7 \pi^4}{120} T^4 + ...
\end{equation}
The last term is the leading contribution in an expansion for $T/M >> 1$.  
The unregularized $J^{med}_{-1}$  has then the same large $T$ dependence $\sim T^4$ as the Polyakov loop potential $\mathcal {U}$ and can mingle with its asymptotics. It is worthwhile pointing out that the complete $PV$ regularized expression, including the $C(T,\mu)$ term has also a $T^4$ dependence at large $T$. However in this case the dependence is due alone to the $C(T,\mu)$ term which, as mentioned before, is  independent of $\phi=\overline{\phi}$ and is thus decoupled from the large $T$ asymptotics of the Polyakov loop degrees of freedom (see also the Appendix). 
A further remark is in order. The gap equations must be solved together with the SPA equations (\ref{StaEq}). Since the  $J_{0}(M_f^2,T,\mu,\phi,\overline{\phi} )$ integrals vanish as $T\rightarrow \infty$, we obtain that $M_f-m_f=0$. This guarantees that in the large $T$ asymptotics the explicit chiral symmetry breaking pattern of the model is recovered. We emphasize once again that this result can only be obtained if the regulator is kept in Dirac and Fermi contributions of the one-loop quark integrals. It is easy to understand that failing to do so, the matter parts will diverge at large $T$, driving instead the solutions of the gap equations to $M_f\rightarrow 0$. The same argument as above in (\ref{int4}) can be applied to infer that the not regularized $J^{med}_{0}$ diverges as $T^2$. Details are discussed in the Appendix, part B.
We shall return to these large $T$ asymptotic features in our numerical results presented in the figures below.  

\section {Numerical results}

Before discussing the numerical results we specify in Table \ref{PolyakovPot} several types of Polyakov potentials proposed in the literature which we shall employ in our analysis. 
The Polyakov potential and the pure gauge QCD Lagrangian should share the $Z(3)$ center symmetry, spontaneously broken above a critical temperature (for $T\rightarrow \infty$ the minimum should correspond to $\left|\phi\right|=1$).
The polynomial form $\mathcal{U}^{I}$  is based on a Ginzburg-Landau ansatz using the terms with the required symmetry (the cubic $\phi$ terms are needed to break the $U(1)$ symmetry down to the $Z(3)$), the exponential term in $\mathcal{U}^{III}$ was derived in the strong coupling expansion of the lattice QCD action and the logarithmic term (which appears in $\mathcal{U}^{II}$, $\mathcal{U}^{III}$ and $\mathcal{U}^{IV}$) is inspired by the $SU(3)$ Haar measure of group integration.

In Table \ref{ParamSets} the parameters for the NJLH (sets I \cite{Osipov:2004}, II \cite{Hiller:2010}) and NJLH8 model Lagrangians (sets III \cite{Hatsuda:1994}, IV\cite{Bhattacharyya:2010}) are collected, obtained by fitting low lying meson characteristics in the vacuum.  Sets I and III have been regularized with PV, the remaining with a sharp 3D cutoff, we denote by $\Lambda$ generically the PV regulating mass and the 3D cutoff in the respective sets. The remaining parameters are defined in the caption of the Table. All parameters are left unchanged at finite $T$ and $\mu$, except that in some examples discussed below the regulator will be removed from the finite medium contributions.

\begin{table}[htb]\scriptsize
\begin{center}
\begin{tabular}[c]{|l|l|}
\hline
\mbox{\normalsize{Polyakov potential}} & \mbox{\normalsize{Parameters}}\\
\hline\hline
\mbox{\tiny{
$\begin{aligned}
\frac{\mathcal{U}^{I}\left(\phi,\overline{\phi},T\right)}{T^4}&=-\frac{b_2\left(T\right)}{2}\overline{\phi}\phi-\frac{b_3}{6}\left(\phi^3+\overline{\phi}^3\right)+\frac{b_4}{4}\left(\overline{\phi}\phi\right)^2\\
b_2(T)&=a_0+a_1 \frac{T_0}{T}+a_2 \left(\frac{T_0}{T}\right)^2+a_3 \left(\frac{T_0}{T}\right)^3
\end{aligned}
$
}}
&
\begin{tabular}{cccc}
$a_0$ & $a_1$ & $a_2$ & $a_3$ \\
$6.75$ & $-1.95$ & $2.625$ & $7.44$\\
 $b_3$ & $b_4$ &&\\
$0.75$ & $7.5$ &&
\end{tabular}
 \\
\hline
\mbox{\tiny{
$
\begin{aligned}
\frac{\mathcal{U}^{II}\left(\phi,\overline{\phi},T\right)}{T^4}&=-\frac{b_2\left(T\right)}{2}\overline{\phi}\phi
+b_3\left(\frac{T_0}{T}\right)^3\ln\left[B\left(\phi,\overline{\phi}\right)\right]\\
B\left(\phi,\overline{\phi}\right)&=1-6\overline{\phi}\phi+4\left(\phi^3+\overline{\phi}^3\right)-3\left(\overline{\phi}\phi\right)^2
\end{aligned}
$
}}

&
\begin{tabular}{ccccc}
$a_0$ & $a_1$ & $a_2$ & $a_3$\\
$3.51$ & $-2.47$ & $15.2$ & $0$ \\
$b_3$ & & & &\\ 
$-1.75$ & & & &
\end{tabular}
\\
\hline
\mbox{\tiny{
$\begin{aligned}
\frac{\mathcal{U}^{III}\left(\phi,\overline{\phi},T\right)}{T^4}=-\frac{b}{T^3}\left(54 e^{-\frac{a}{T}}\overline{\phi}\phi+\ln\left[B\left(\phi,\overline{\phi}\right)\right]\right)
\end{aligned}
$
}}
&
\begin{tabular}{cc}
$a$ & $b$\\
$664~\mathrm{MeV}$ & $0.03\Lambda^3$
\end{tabular}
\\
\hline
\mbox{\tiny{
$
\begin{aligned}
\frac{\mathcal{U}^{IV}\left(\phi,\overline{\phi},T\right)}{T^4}=&-\frac{b_2\left(T\right)}{2}\overline{\phi}\phi-\frac{b_3}{6}\left(\phi^3+\overline{\phi}^3\right)+\frac{b_4}{4}\left(\overline{\phi}\phi\right)^2\\
&-K \ln[\frac{27}{24 \pi^2}B\left(\phi,\overline{\phi}\right)]
\end{aligned}
$
}}
&
\begin{tabular}{cccc}
$a_0$ & $a_1$ & $a_2$ & $a_3$\\
$6.75$ & $-1.95$ & $2.625$ & $-7.44$\\
$b_3$ & $b_4$ &   & \\
 $0.75$ & $7.5$ &   & \\
\end{tabular}
\\
\hline
\end{tabular}
\end{center}
\caption{Some of the proposed forms for the Polyakov potential: $\mathcal{U}^{I}$ from \cite{Weise:2006}, $\mathcal{U}^{II}$ from \cite{Roessner:2006}, $\mathcal{U}^{III}$ from \cite{Fukushima:2008}
 and $\mathcal{U}^{IV}$ from \cite{Bhattacharyya:2010}. Fits to the pure gauge lattice-QCD data result in $T_0=270~\mathrm{MeV}$, however in the presence of dynamical quarks a rescaling is usually applied and a lower value for this parameter (which appears in $\mathcal{U}^{I/II/IV}$) is used. The parameter, $K$, in $\mathcal{U}^{IV}$, depends on the choice of parametrization for the quark interactions (in \cite{Bhattacharyya:2010} it is fitted to reproduce the pressure and transition temperature of lattice data). In \cite{Fukushima:2008} the parameter $b$ was chosen so as to lead to simultaneous confinement and chiral transitions around $T\approx 200 \mathrm{MeV}$ ($b=0.03(631.4\mathrm{MeV})^3$).
In \cite{Hell:2010} $\mathcal{U}^{II}$ is used with a different parametrization.}
\label{PolyakovPot}
\end{table}

\begin{table}[htb]\scriptsize
\begin{tabular}[c]{|l||r|r|r|r|r|r|r| }
\hline
   Sets & $m_u$ & $m_s$ & $\Lambda$ & $G$ & $\kappa$ & $g_1$ & $g_2$ \\
&\begin{small}($\mathrm{MeV}$)\end{small} &\begin{small} ($\mathrm{MeV}$)\end{small} &\begin{small}($\mathrm{MeV}$)\end{small} &\begin{small}($\mathrm{GeV}^{-2}$)\end{small} &\begin{small}($\mathrm{GeV}^{-5}$)\end{small} &\begin{small}($\mathrm{GeV}^{-8}$)\end{small} &\begin{small}($\mathrm{GeV}^{-8}$)\end{small}\\
\hline\hline
I \cite{Osipov:2004} & $5.3$ & $170$ &  $920$ & $8.89$ & $-687$ & $0$ & $0$\\
\hline
II \cite{Hatsuda:1994}& $5.5$ & $135.7$ &  $631.4$ & $9.21$ & $-740.6$ & $0$ & $0$\\
\hline
III \cite{Hiller:2010} & $5.9$ & $186$ &  $851$ & $10.92$ & $-1001$ & $1000$ & $-47$\\
\hline
IV \cite{Bhattacharyya:2010}& $5.5$ & $183.468$ &  $637.720$ & $7.165$ & $-720.245$ & $2193$ & $-589$\\
\hline
\end{tabular}
\caption{Parameter sets I and II are for the NJLH model, using PV and 3D regularizations respectively. Sets III and IV are for the NJLH model extended to include the stabilizing eight quark interactions, with strengths $g_1$ and $g_2$, with PV and 3D regularizations respectively. The remaining parameters are the current quark mass values $m_u,m_s$, the cutoff $\Lambda$ that indicates either the PV regulating mass or the 3D cutoff, the 4q interaction strength $G$, and the 't Hooft interaction strength $\kappa$}
\label{ParamSets}
\end{table}

\begin{figure}[thp]
\begin{center}
\label{PotPol1PNJL}
\subfigure[]{\label{DegPNJL}\includegraphics[width= 0.3 \columnwidth]{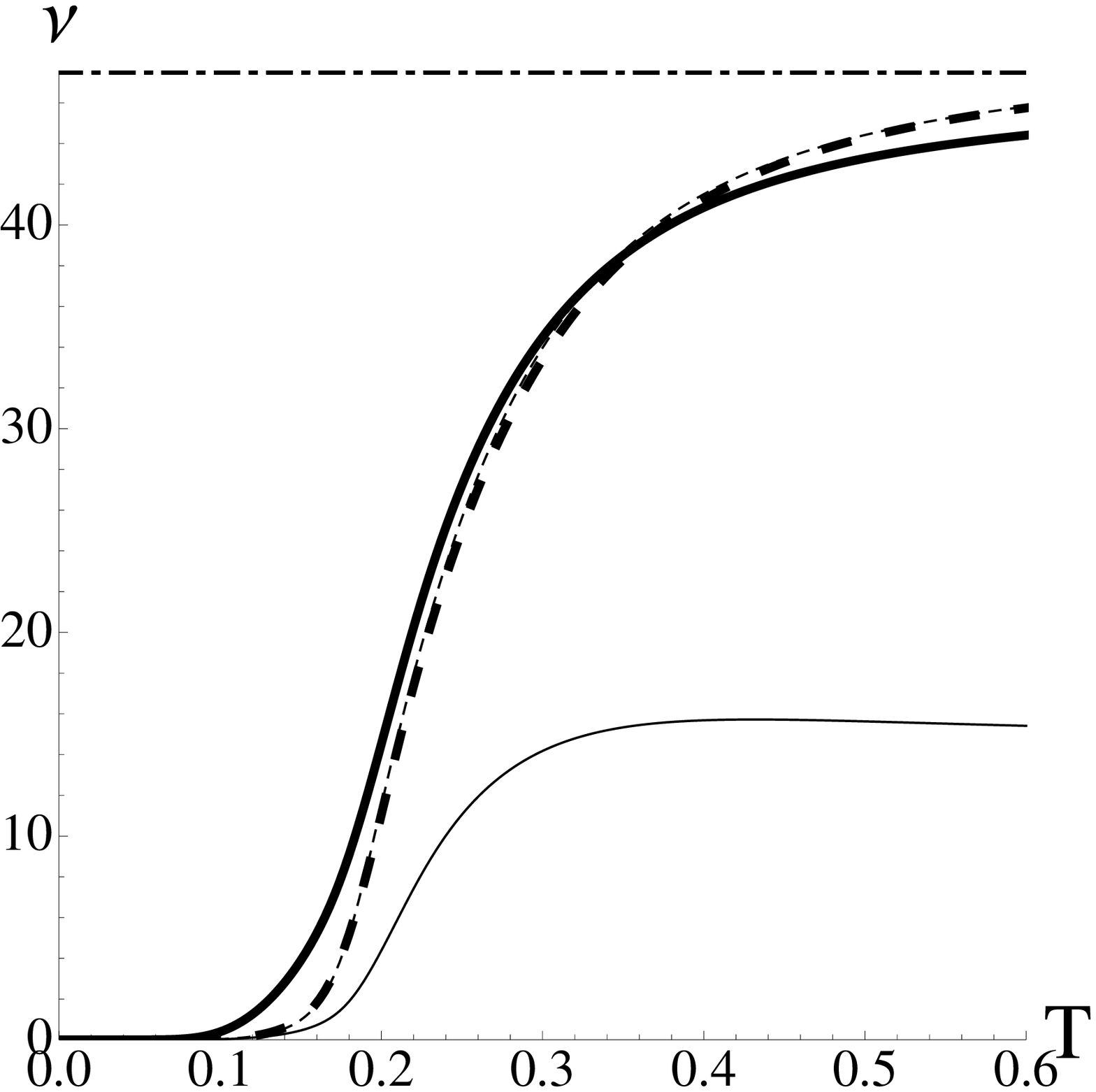}}
\subfigure[]{\label{TesteMi4D}\includegraphics[width= 0.3 \columnwidth]{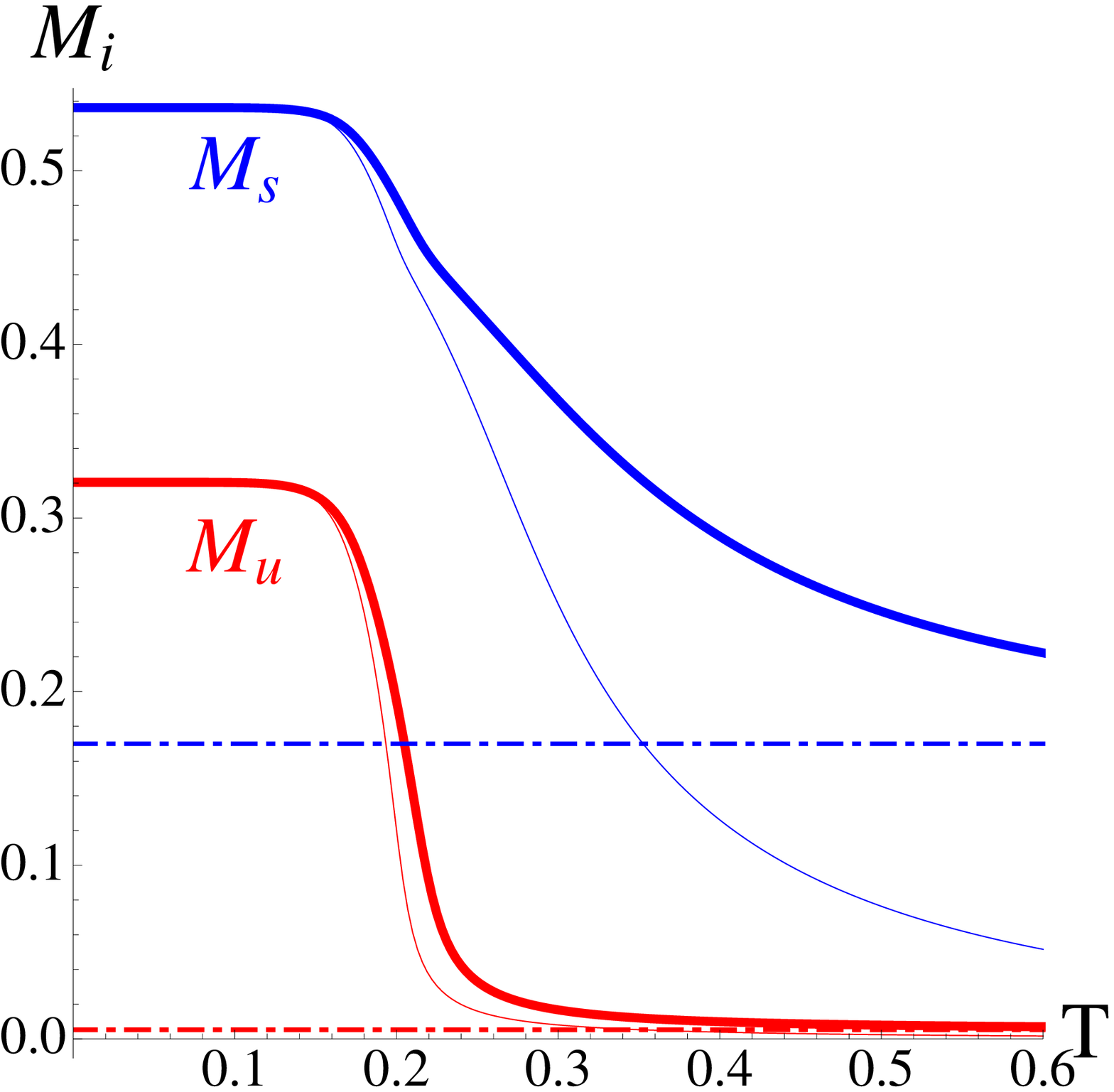}}
\subfigure[]{\label{TesteMi3D}\includegraphics[width= 0.3 \columnwidth]{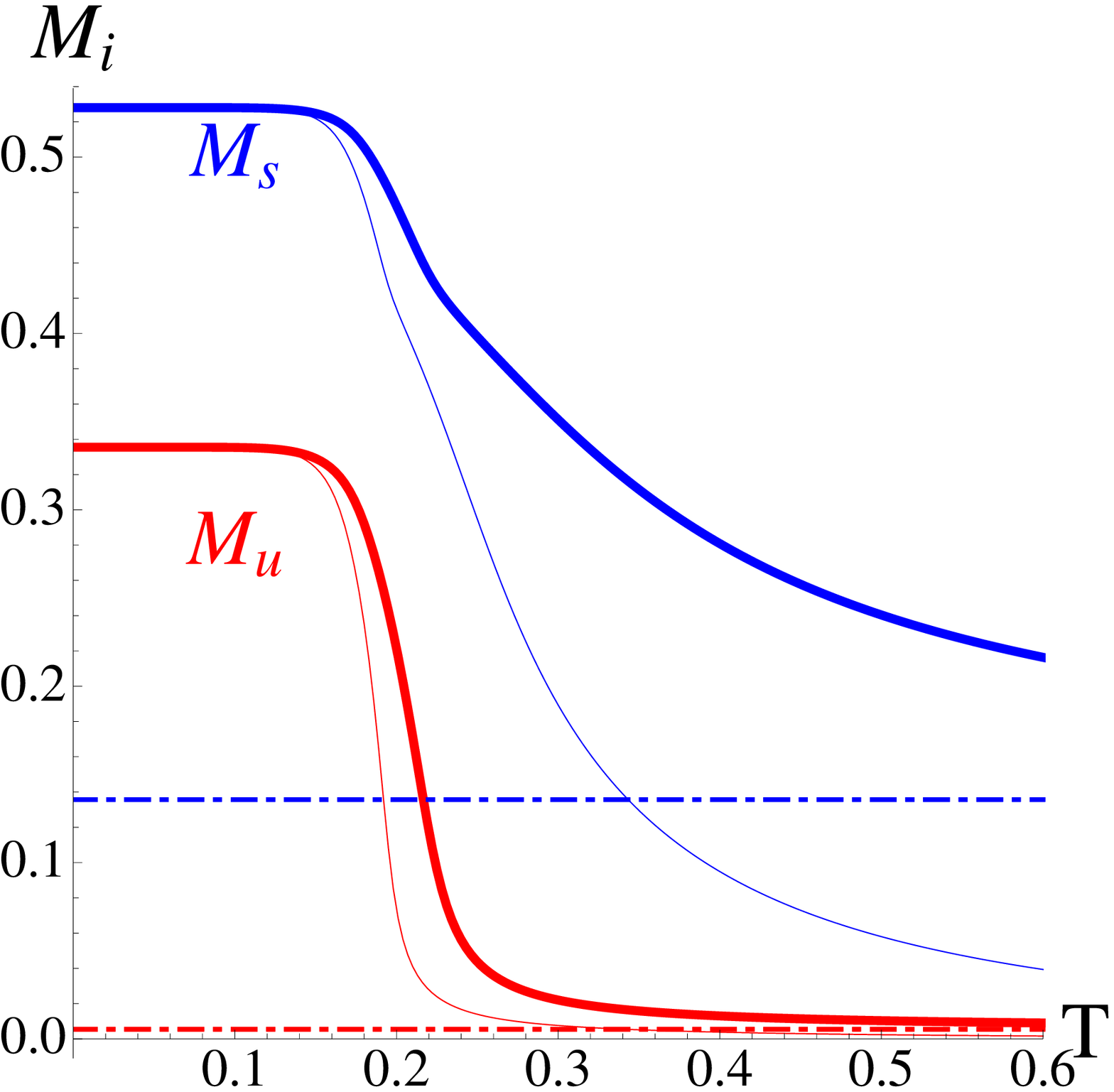}}
\caption{Temperature dependence ($\left[T\right]=\mathrm{GeV}$) at vanishing quark chemical potential  in the PNJLH model with Polyakov potential $\mathcal{U}^{I}$ (see Table \ref{PolyakovPot}, $T_0=0.2~\mathrm{GeV}$) of the following quantities: in Fig.  \ref{DegPNJL} is displayed the number of effective degrees of freedom, $\nu(T)$; the thick solid line corresponds to set I from Table \ref{ParamSets} with PV regularization applied to vacuum and matter part integrals, the thick dashed line is obtained if only the vacuum integrals are PV regularized. Similarly thin lines stand for the degrees of freedom calculated with 3D regularization using set II from Table \ref{ParamSets}, with overall regularization (solid line) and if only the vacuum part is regularized (dashed line); the horizontal dot-dashed line represents the Stefan-Boltzmann limit. In Fig. \ref{TesteMi4D} are shown the dynamical quark masses  of the quarks  ($\left[M_i\right]=\mathrm{GeV}$) corresponding to the set I  
(PV regularization), the thick line refers to the combined use of PV in vacuum and matter parts, the thin line is obtained if only the vacuum part is PV regularized;  the horizontal dot-dashed lines indicate the current quark masses.   In Fig. \ref{TesteMi3D} are displayed the same quantities with the same notation as in Fig. \ref{TesteMi4D} for set II (3D regularization).}
\end{center}
\end{figure}

\begin{figure}[thp]
\begin{center}
\label{PNJLphimufin}
\subfigure[]{\label{PNJLphimu0}\includegraphics[width= 0.3 \columnwidth]{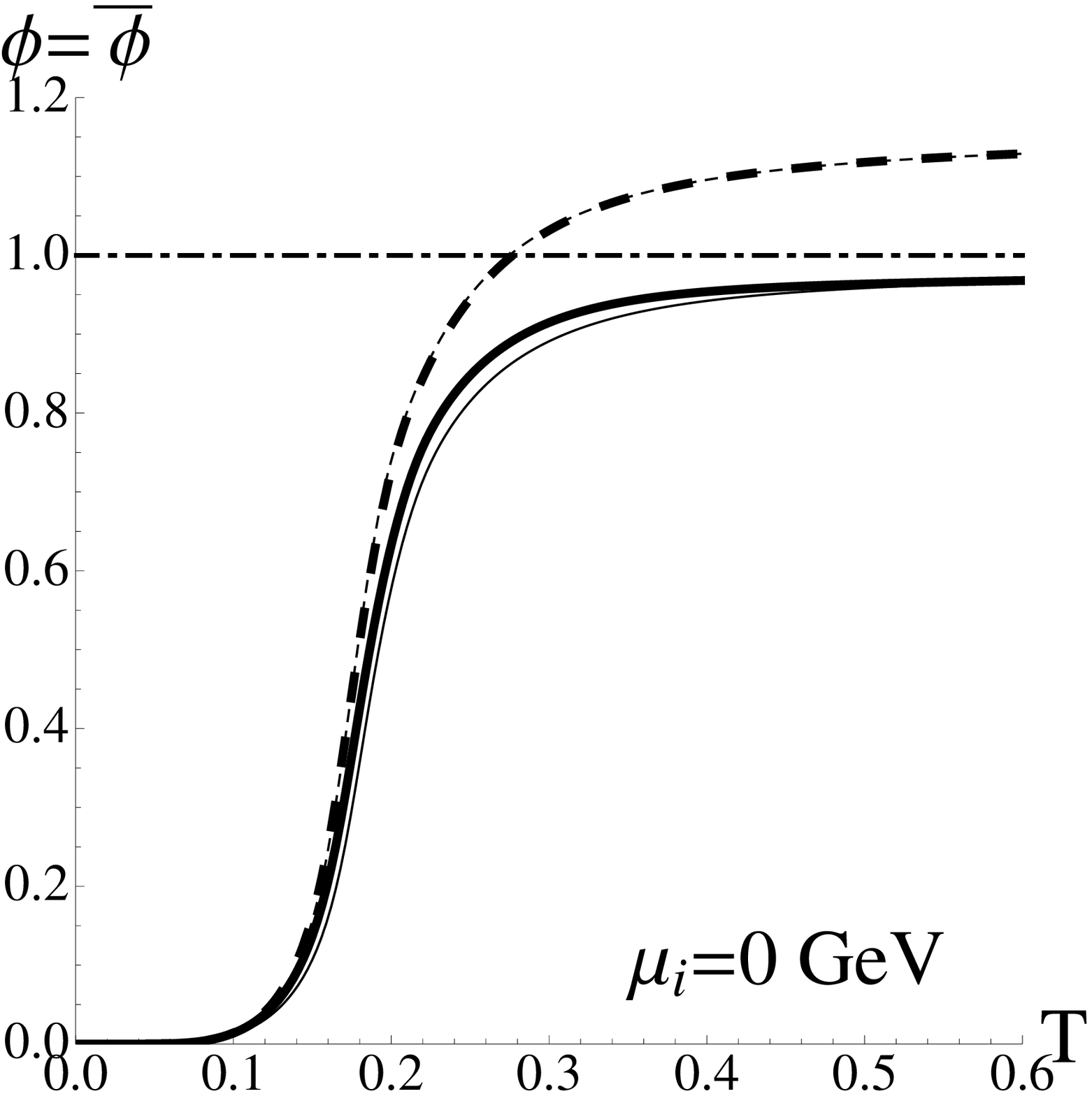}}
\subfigure[]{\label{PNJLphimufin1}\includegraphics[width= 0.3 \columnwidth]{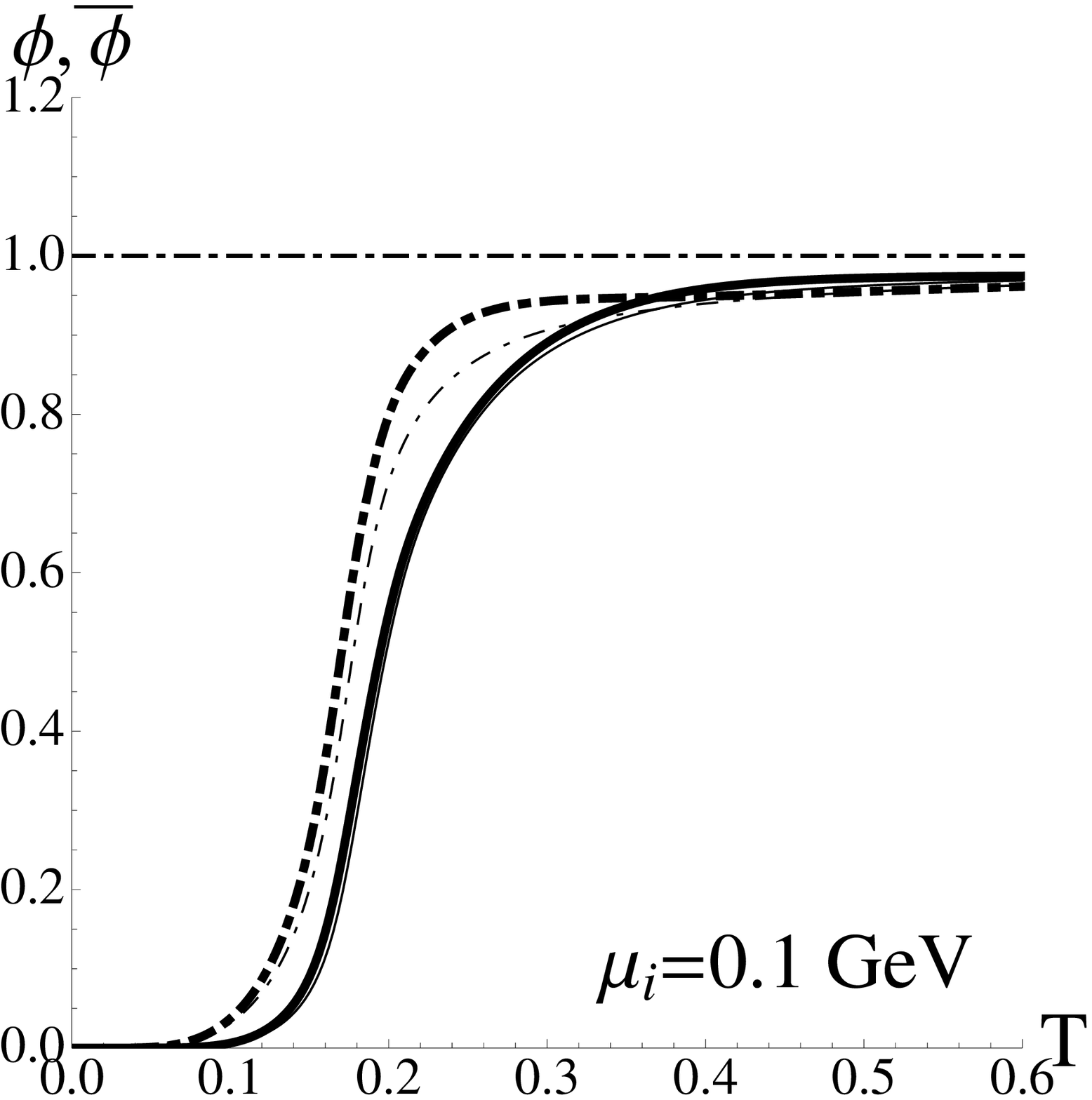}}
\subfigure[]{\label{PNJLphimufin2}\includegraphics[width= 0.3 \columnwidth]{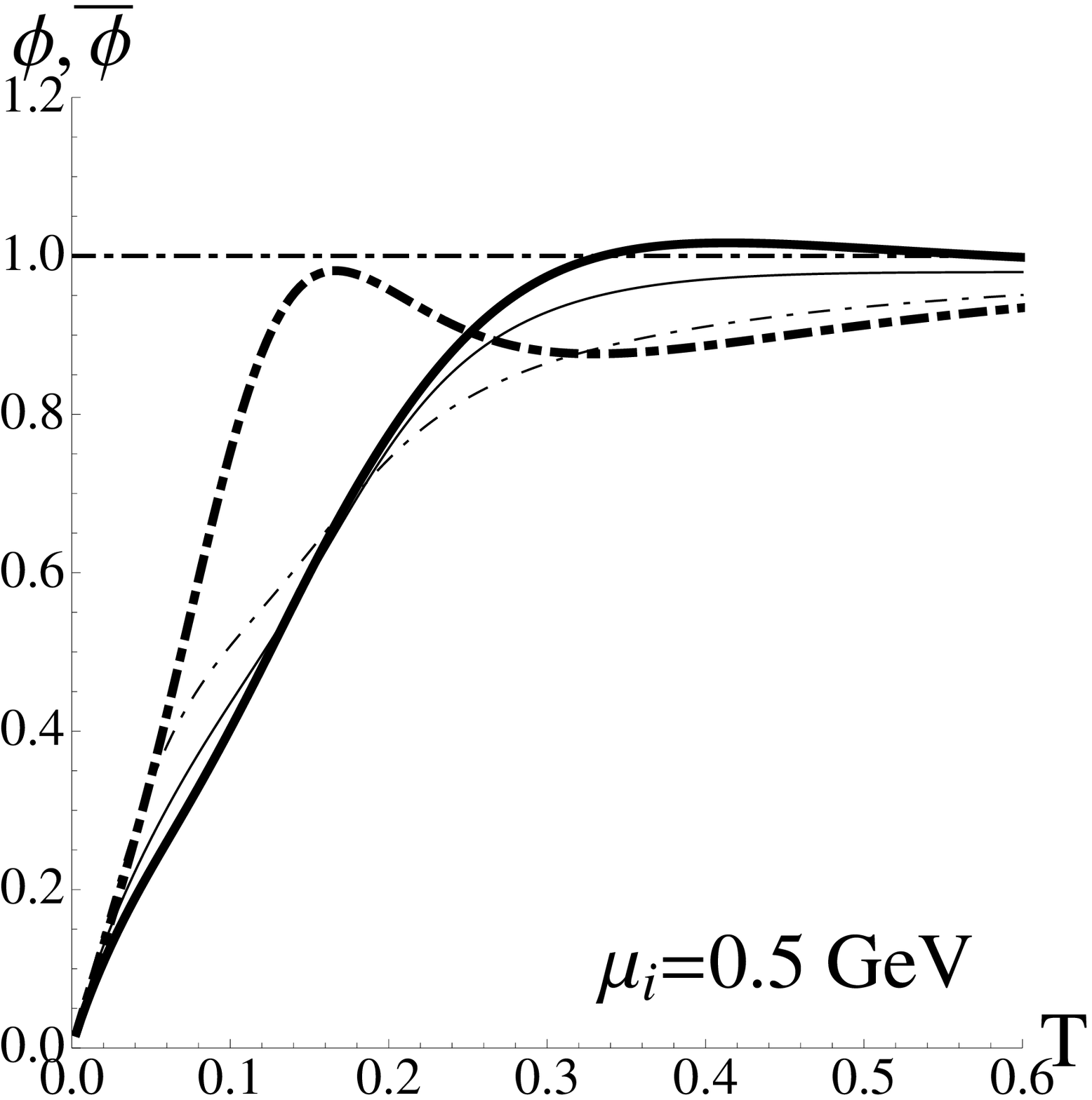}}
\caption{Temperature dependence ($\left[T\right]=\mathrm{GeV}$) of the traced Polyakov loop $\phi$ using the Polyakov potential $\mathcal{U}^{I}$ from Table  \ref{PolyakovPot}, $T_0=0.2~\mathrm{GeV}$  (thick black lines correspond to set (I) (PV regularization) and thin lines to set (II) (3D cutoff) from Table \ref{ParamSets} for the following cases: in \ref{PNJLphimu0} $\mu=0$, the solid lines correspond to the case where regularization is used for all the contributions (vacuum and medium parts) whereas dashed lines correspond to the regularization of the vacuum part only; in \ref{PNJLphimufin1} and \ref{PNJLphimufin2} $\mu \ne 0$ and all curves are calculated regularizing both vacuum and matter integrals, thick refers as before to PV regularization and thin to 3D, and the dot-dashed lines describe the $T$ dependence of $\overline{\phi}$ (which for finite chemical potential no longer coincides with $\phi$); the horizontal dot-dashed line represents the desired asymptotic value for the traced Polyakov loop.}
\end{center}
\end{figure}

Our results are presented in the Figs. 1-5. 
Fig. 1a displays the number of effective degrees of freedom $\nu$  calculated for sets I  and II of Table 2, in presence of the Polyakov loop. As mentioned before, the PV regularization (full thick black line) leads to the correct large $T$ asymptotics, $\nu(T=\infty)=47.5$ (dot-dashed horizontal line) determined as the sum of quark degrees of freedom, (31.5 for 3 flavors) and Polyakov loop degrees of freedom ($16=2\times 8$, polarization$\times$ color) \footnote{Note that, as pointed out in \cite{Fukushima:2008} this is not the case with $\mathcal{U}^{III}$.}
. If one removes the regulator in the finite matter parts, one obtains the dashed thick black line. As opposed to this, the $3D$ sharp cutoff shows a large difference between the full regularized expression (full thin curve) and the one where only the vacuum part is regularized (thin dashed line, which is almost indistinguishable from the thick dashed line). It is worth noting that in the case of the $3D$ regularization, with or without the Polyakov loop,  the quark degrees of freedom are largely underestimated above the crossover temperature if all integrals are regularized, and only if the finite contributions are not, does one obtain the right order of magnitude. The PV regularization fully resolves these problems. The desired approximation to the Stefan-Boltzmann limit is obtained by consistently keeping the regulator in all contributions and most importantly by the implementation of the $C(T,\mu)$ term (\ref{FTP}) in the thermodynamic potential.
  
In Figs.  \ref{TesteMi4D} and \ref{TesteMi3D}   the constituent quark mass profiles are shown as functions of $T$ for sets I and II, according to the different regularizations employed, PV and 3D respectively. In \cite{PCosta:2010} it has been noticed that the $3D$ regularization applied to the vacuum part only, leads to a sharp fall-off below the current quark mass values, Fig. 1c, thin lines. This feature is also present if the PV regularization is applied to the vacuum part only, Fig. 1b, thin lines. The reason was discussed above and related with thermodynamic inconsistency if the matter parts are not regularized.  
The full lines in both Figures correspond to regularizing also the finite matter parts. Only then does one recover in the large $T$ limit the explicit breaking pattern of chiral symmetry, in both the PV and 3D regularizations.

In Fig. 2 we address another issue which is connected with the regularization procedure as well.
As mentioned earlier, the choice of using a 3D cutoff only in the vacuum integrals was
necessary to reproduce the correct behavior of thermodynamical quantities in the NJL model without the Polyakov loop \cite{Klevansky:1994}. When applied to the PNJL model using the polynomial form of the Polyakov loop potential, this procedure results in an asymptotic value for $\phi$ larger than unity when $T\rightarrow\infty$ (as well as negative susceptibilities) \cite{Sazaki:2007}.  This can be seen in Figure \ref{PNJLphimu0},  in the case of the PNJLH model, for vanishing chemical potential. The thick and thin dashed lines correspond to the removal of the regulator in the finite parts in the PV and 3D cases respectively. This undesired result is presented as one of the reasons behind the adoption of ``improved'' forms of the potential which include a logarithmic term which diverges as $\phi,\overline{\phi}\rightarrow 1$ thus resulting in the correct asymptotic behavior. However we stress that if the matter parts are also regularized,  (the solid thick and thin lines in \ref{PNJLphimu0}) the correct asymptotics is obtained.
In Figs. (2b) and (2c) we show $\phi,\overline{\phi}$ (full and dot-dashed lines respectively) if a finite chemical potential is introduced, and with all integrals regularized, for the PV (thick lines) and 3D (thin lines). Although large differences are observed due to the two regularizations at finite $T$, the asymptotic values will tend all to unity. 

\begin{figure}[htp]
\begin{center}
\label{PotLogPNJL}
\subfigure[]{\label{PNJLphimu0Log}\includegraphics[width= 0.3 \columnwidth]{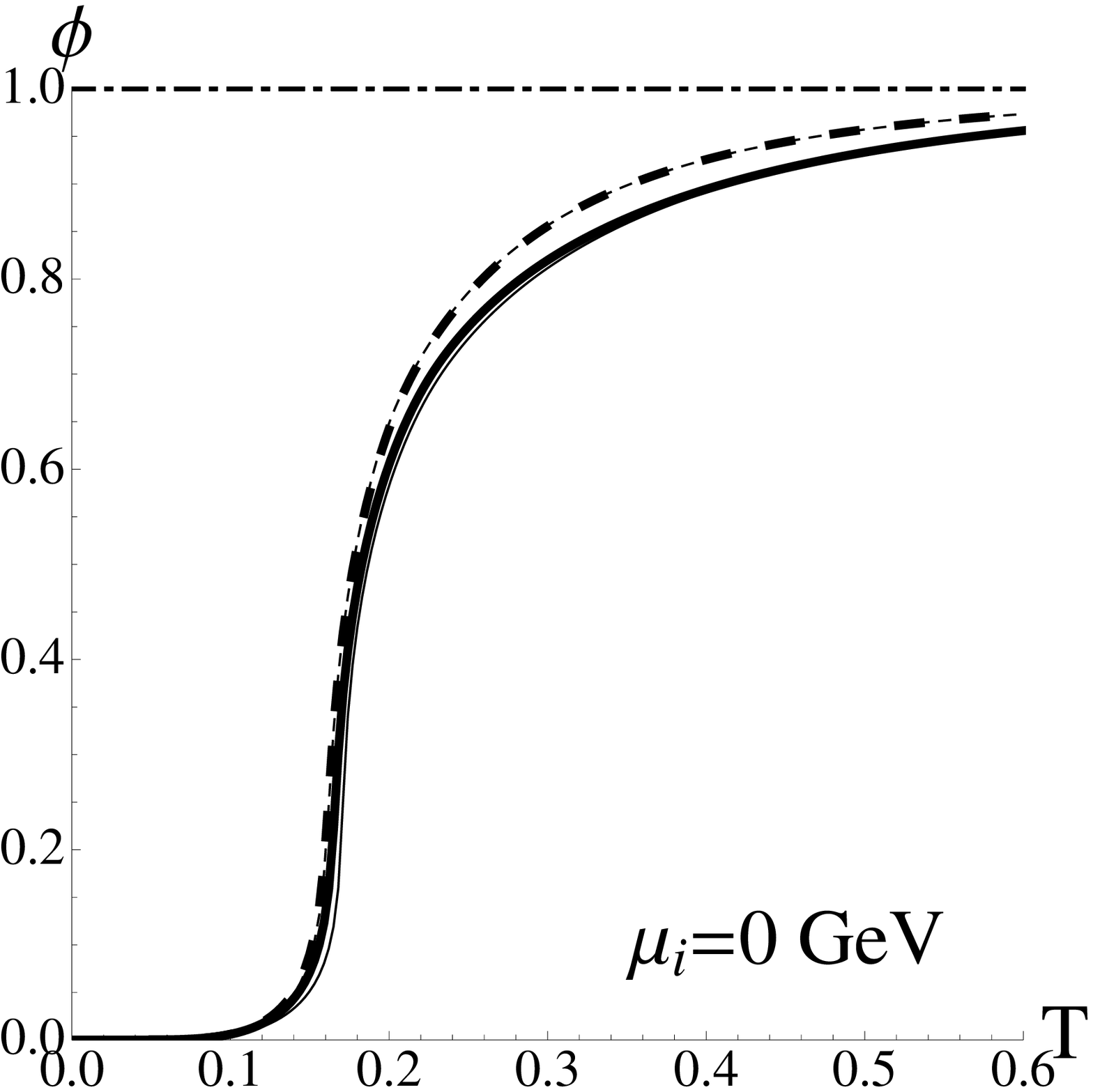}}
\subfigure[]{\label{TesteMi4DLog}\includegraphics[width= 0.3 \columnwidth]{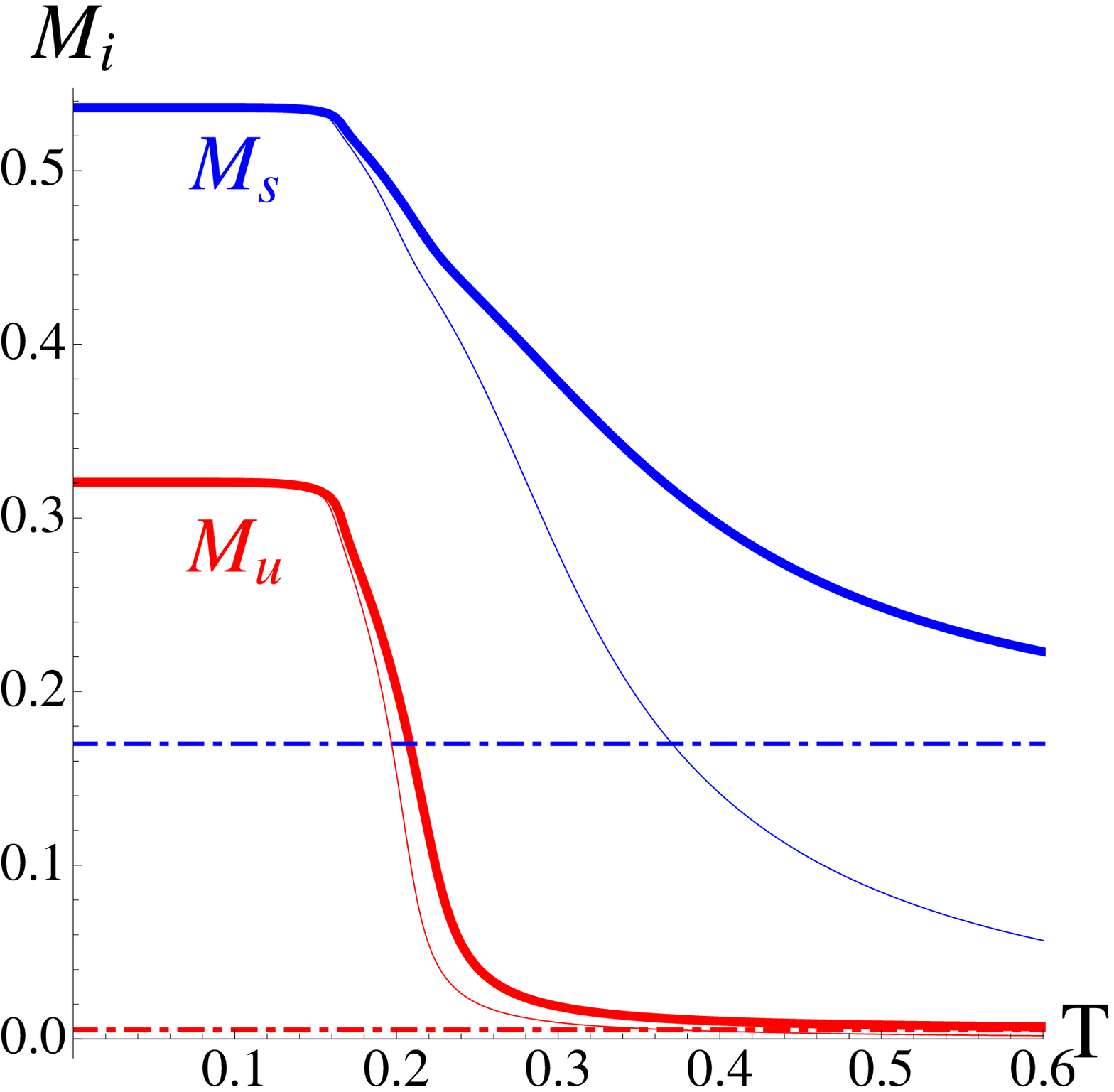}}
\subfigure[]{\label{TesteMi3DLog}\includegraphics[width= 0.3 \columnwidth]{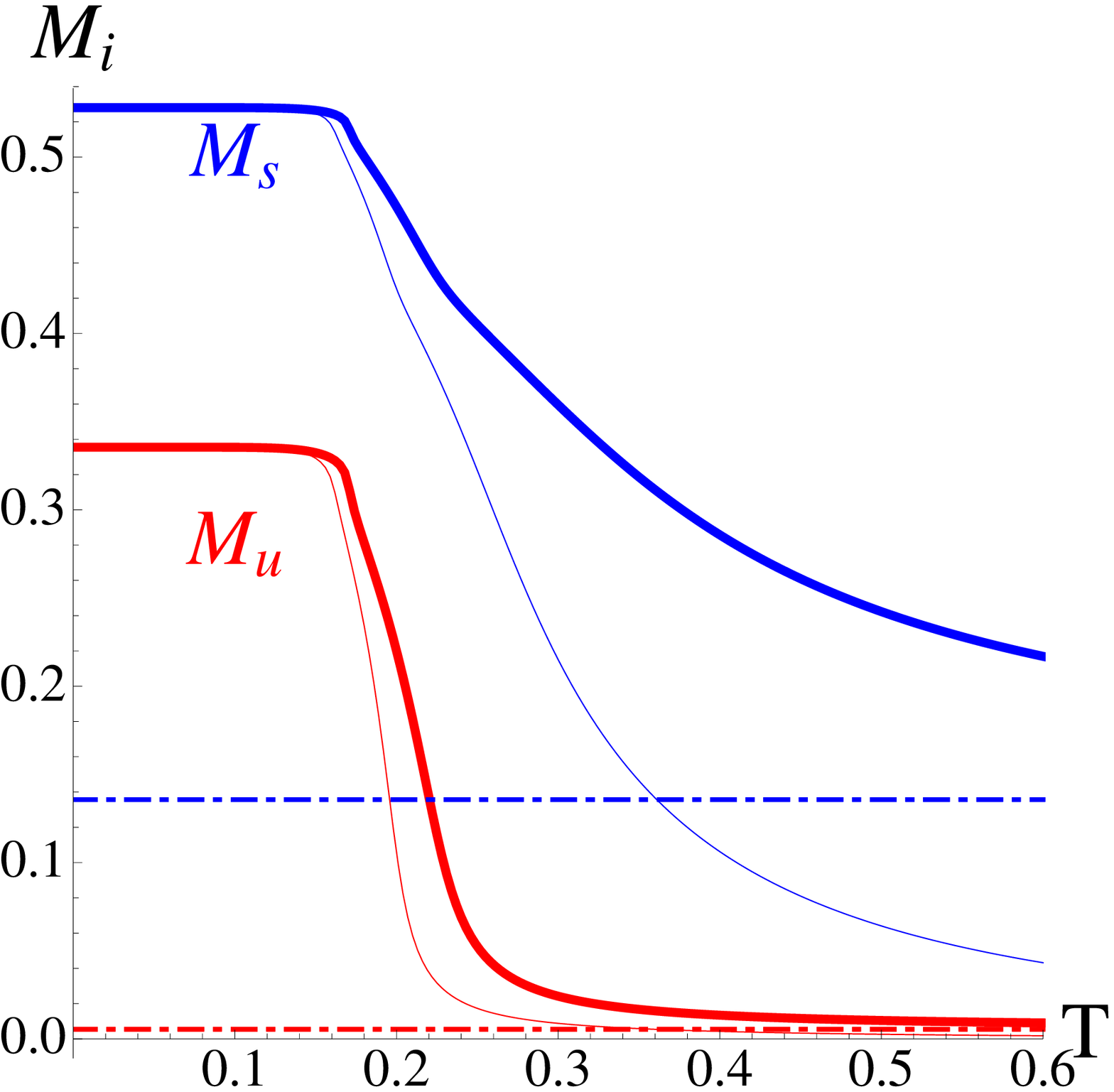}}
\caption{
The temperature dependence ($\left[T\right]=\mathrm{GeV}$, $\mu_i=0$) of the traced Polyakov loop $\phi$ is represented using the Polyakov potential $\mathcal{U}^{II}$ from Table \ref{PolyakovPot}, $T_0=0.200~\mathrm{GeV}$, thick solid lines refer to the parameter set (I) from Table \ref{ParamSets} (overall PV regularization), thin solid lines are for set (II)  in \ref{PNJLphimu0Log} (overall 3D regularization, they are almost coincident; the dashed curves correspond to the regularization of the corresponding vacuum parts only, also almost indistinguishable; the horizontal dot-dashed line represents the expected asymptotic value for the traced Polyakov loop. In Figs. \ref{TesteMi4DLog} and \ref{TesteMi3DLog} are displayed in thick lines the corresponding dynamical masses of the quarks ($\left[M_i\right]=\mathrm{GeV}$), the middle panel for PV regularization, the right panel for 3D cutoff, in both cases the current quark mass values (dot-dashed horizontal lines) are approached asymptotically; the thin lines denote the cases where only the vacuum parts are regularized, they approach instead zero asymptotically.}

\end{center}
\end{figure}

\begin{figure}[htp]
\begin{center}.
\label{PotLog2PNJL8q}
\subfigure[]{\label{PNJLphimu0Log2}\includegraphics[width= 0.3 \columnwidth]{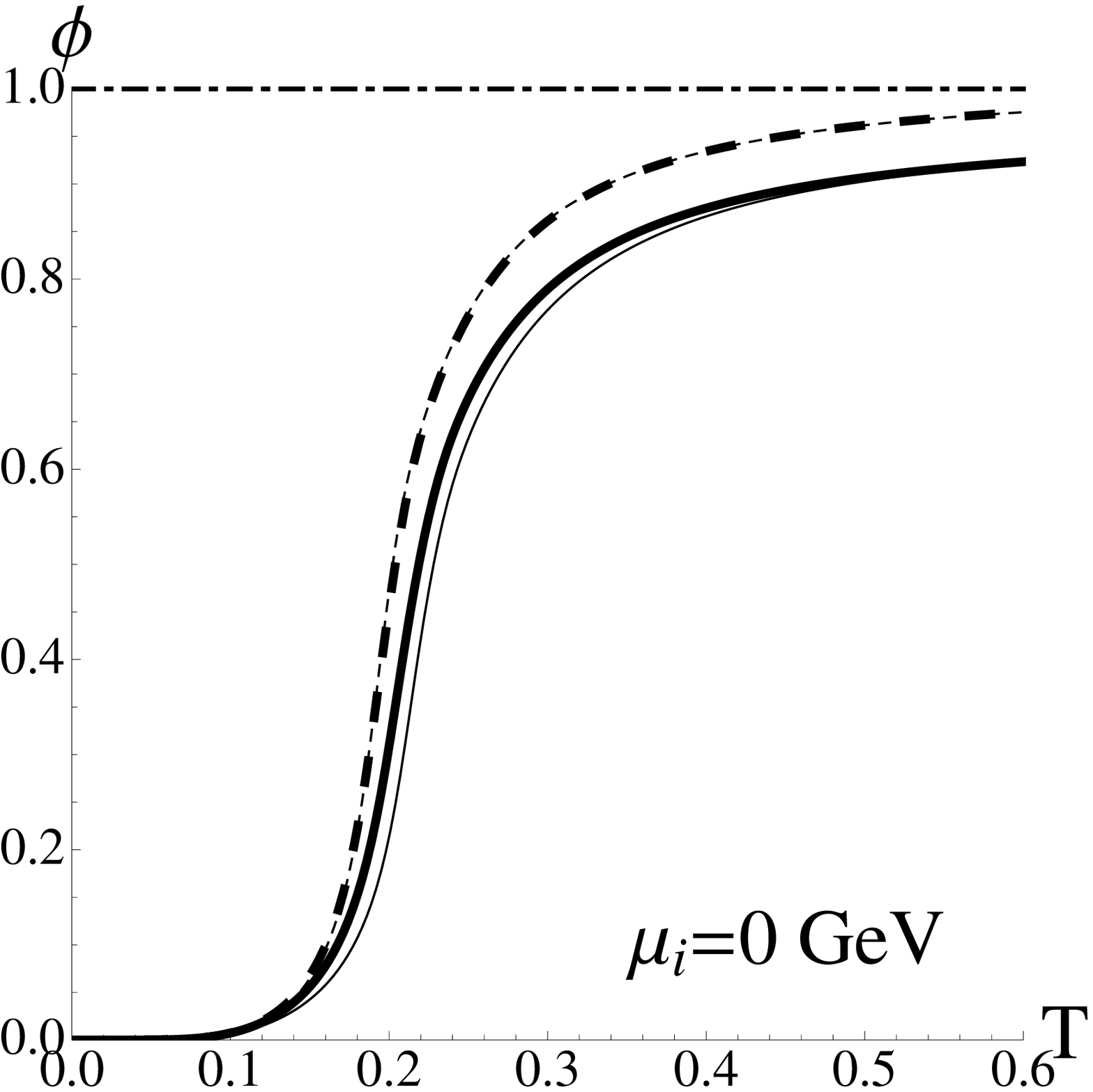}}
\subfigure[]{\label{TesteMi4DLog2}\includegraphics[width= 0.3 \columnwidth]{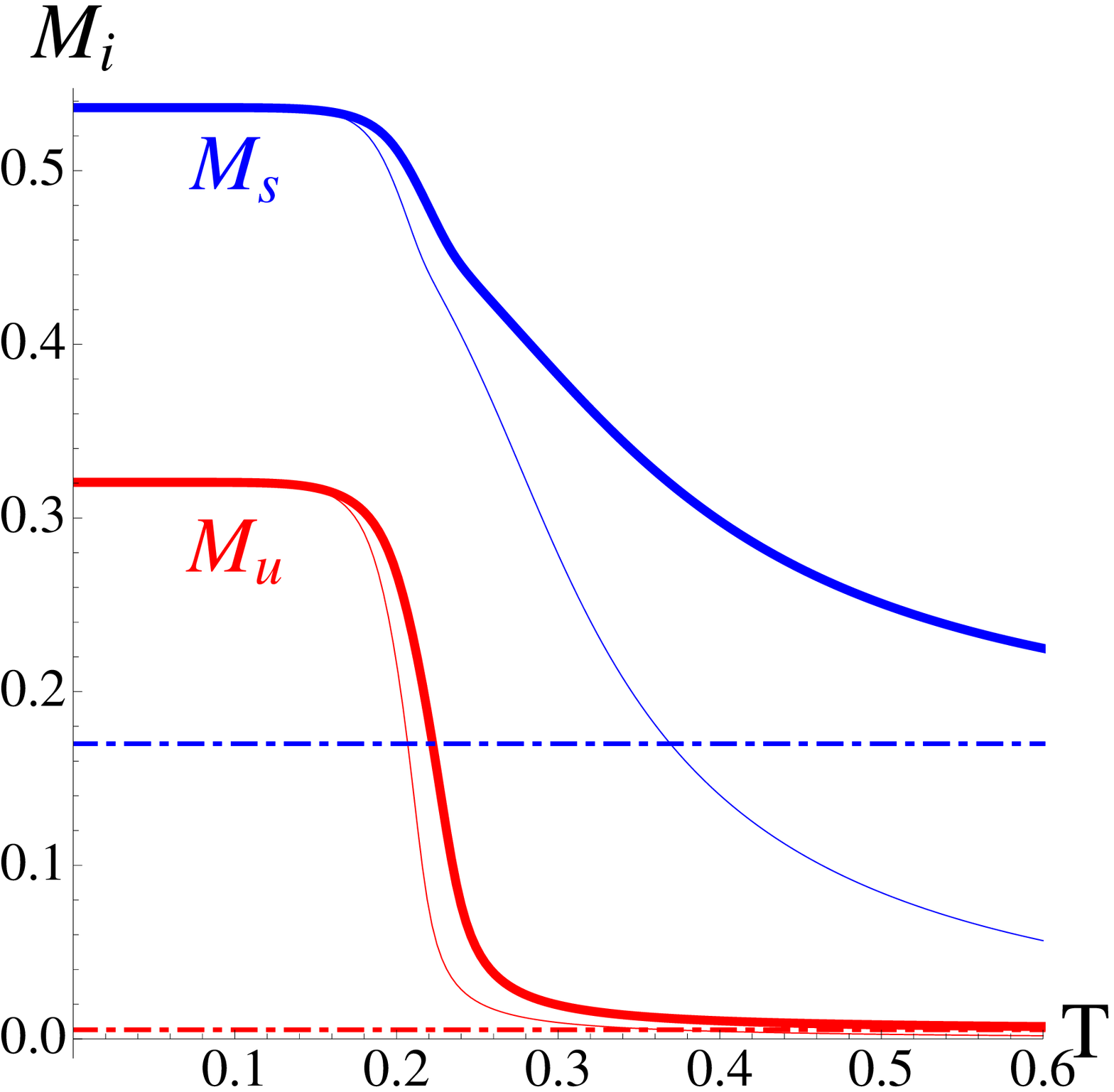}}
\subfigure[]{\label{TesteMi3DLog2}\includegraphics[width= 0.3 \columnwidth]{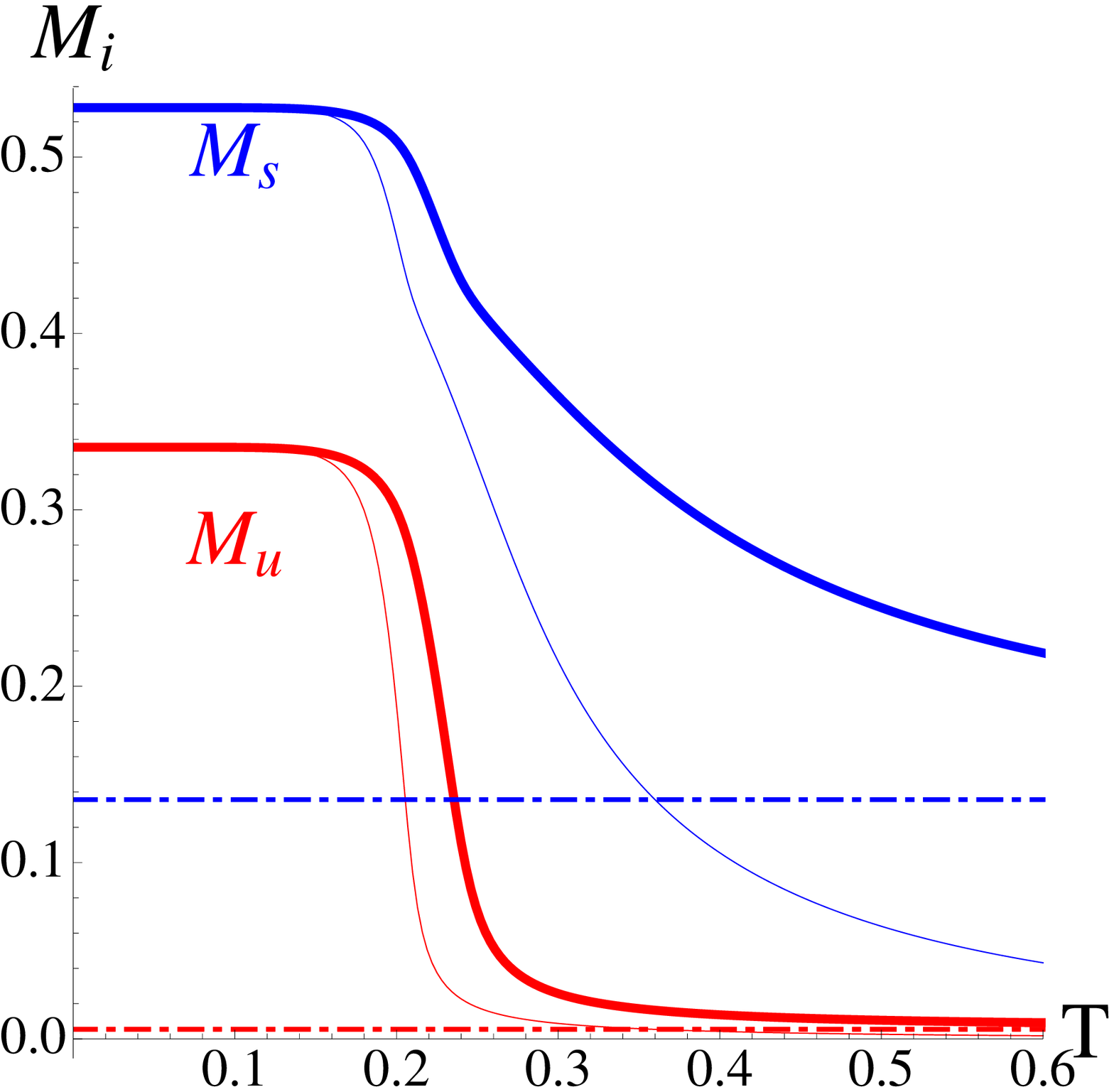}}
\caption{
Same as in the previous Figure (same notation for the curves) using the potential $\mathcal{U}^{III}$ (see Table \ref{PolyakovPot}, $T_0=0.27~\mathrm{GeV}$): thick lines using set I from Table \ref{ParamSets} and thin lines  using set II from Table \ref{ParamSets}
.}
\end{center}
\end{figure}

In the Figs. \ref{PNJLphimu0Log} and \ref{PNJLphimu0Log2} one sees that 
the introduction of a logarithmic term in the Polyakov potential  restricts the value of the traced Polyakov loop to below unity independently of the choice of regularization and whether matter parts are regularized or not, ensuring the correct asymptotics. However this does not change the conclusion that the correct asymptotic values of the constituent quark masses
can only be obtained if the finite integrals are regularized (see Figs. \ref{TesteMi4DLog}, \ref{TesteMi3DLog}, \ref{TesteMi4DLog2} and \ref{TesteMi3DLog2}).

Finally in Fig. 5 we show that the effect of inclusion of the $8q$ Lagrangian for weak coupling strengths $g_1$ and 
$g_2$ does not alter the main results discussed so far. This result corroborates our findings \cite{Osipov:2007b}
\cite{Hiller:2010} that as long as spontaneous breaking of chiral symmetry is induced by the $4q$ coupling strength $G$, the thermal results of NJLH and NJLH8 models will not lead to significant differences. The inclusion of the Polyakov physics does not change this expectation. The parameter set IV is taken from \cite{Bhattacharyya:2010}, where the Polyakov loop was considered together with the NJLH8 Lagrangian, and the 3D regularization employed. Note that in this case the minimum of the Polyakov potential 
$\mathcal{U}^{IV}$ at large $T$ is at $\phi(T\rightarrow \infty)=0.878$ and not unity.  As discussed before the thermodynamical asymptotics for $\phi$ coincides with the minimum of the Polyakov potential if the matter parts are regularized. That's what happens in this case as well (solid lines), as opposed to the case where the regulator is removed (dashed lines). The parameter set III taken from \cite{Hiller:2010} was chosen also in the weak $8q$ coupling limit for comparison when the PV is used instead.  
\begin{figure}[htp]
\begin{center}
\label{PotLog3PNJL8q}
\subfigure[]{\label{PNJLphimu0Ind}\includegraphics[width= 0.3 \columnwidth]{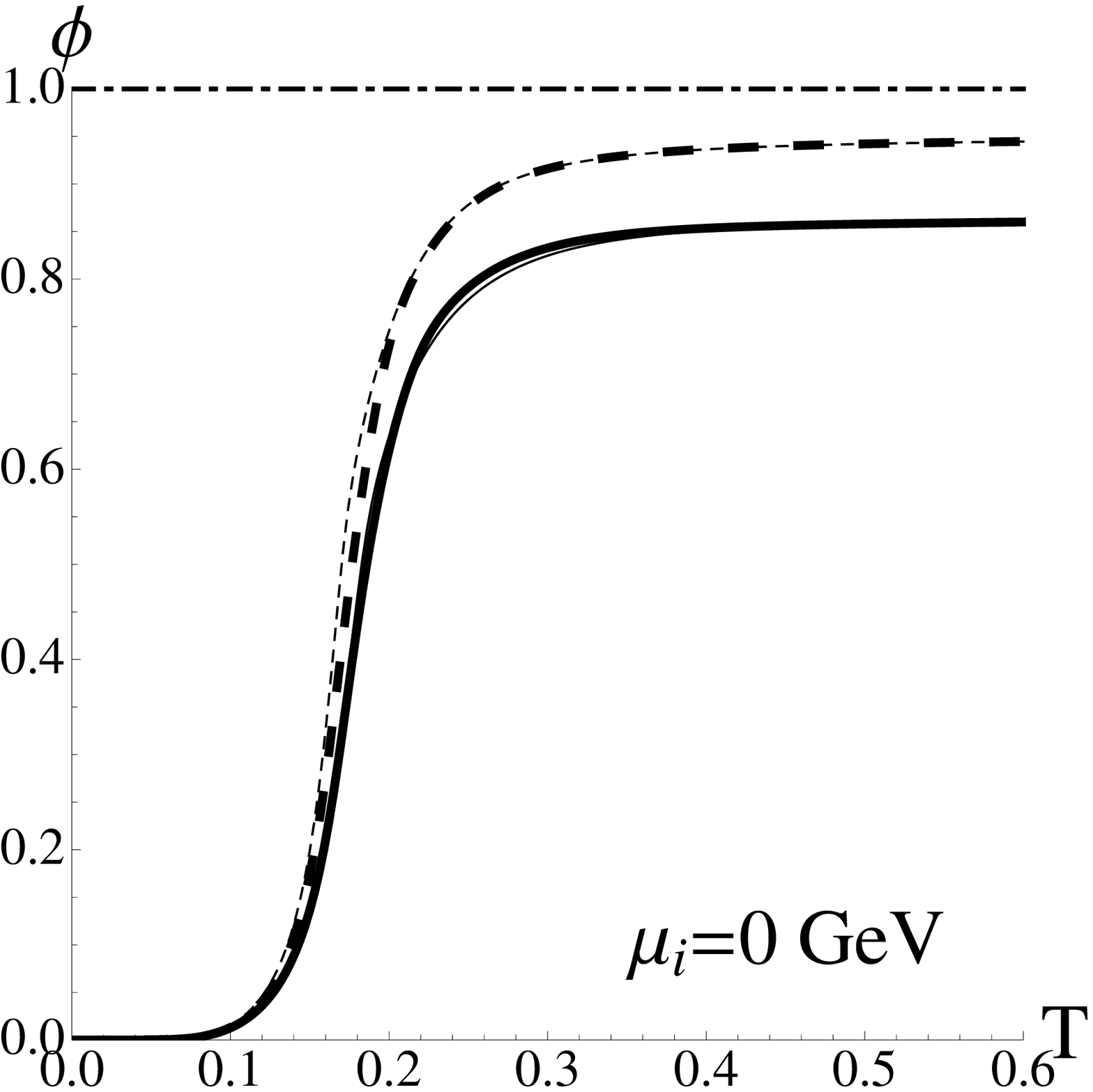}}
\subfigure[]{\label{TesteMi4DInd}\includegraphics[width= 0.3 \columnwidth]{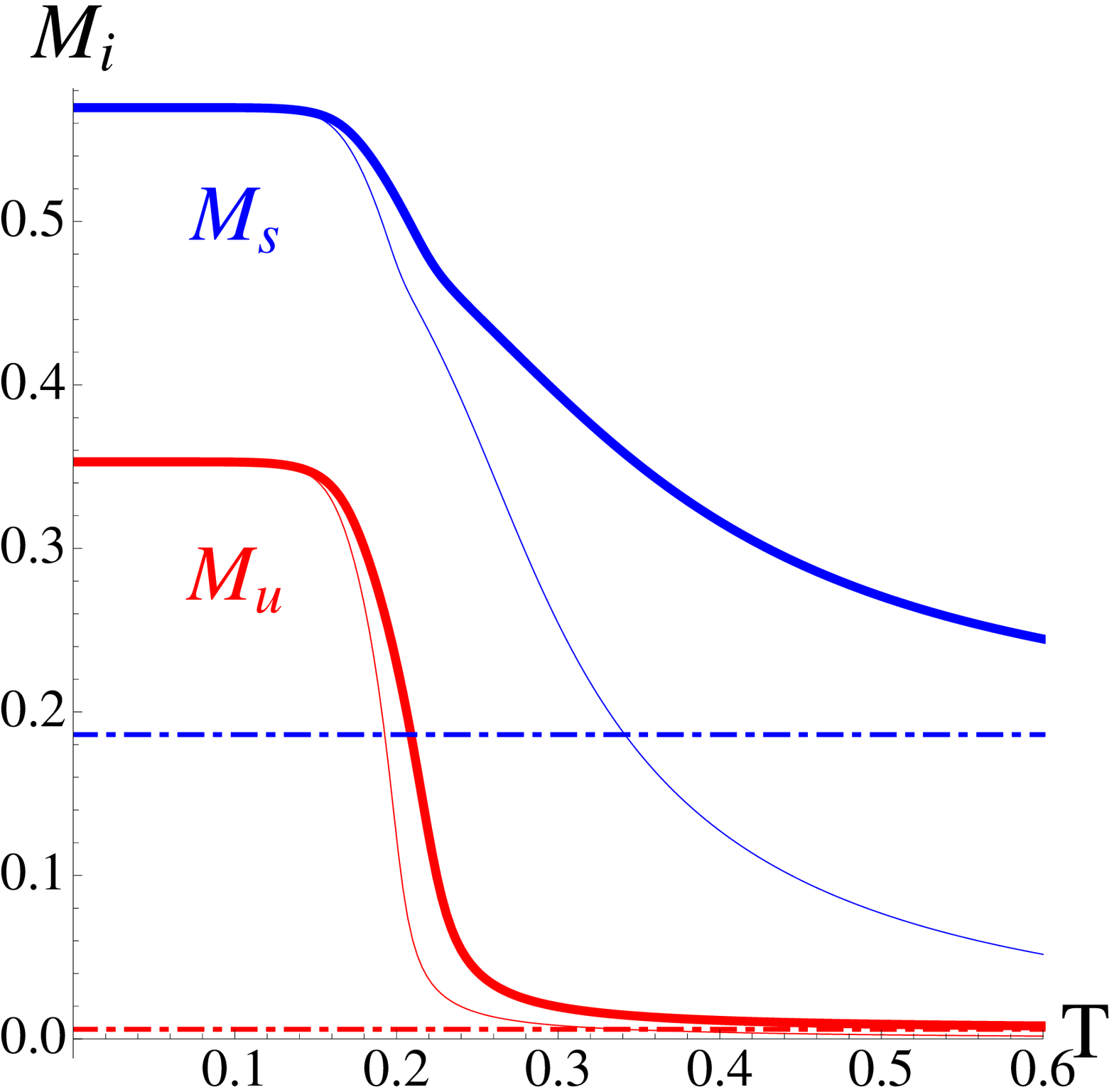}}
\subfigure[]{\label{TesteMi3DInd}\includegraphics[width= 0.3 \columnwidth]{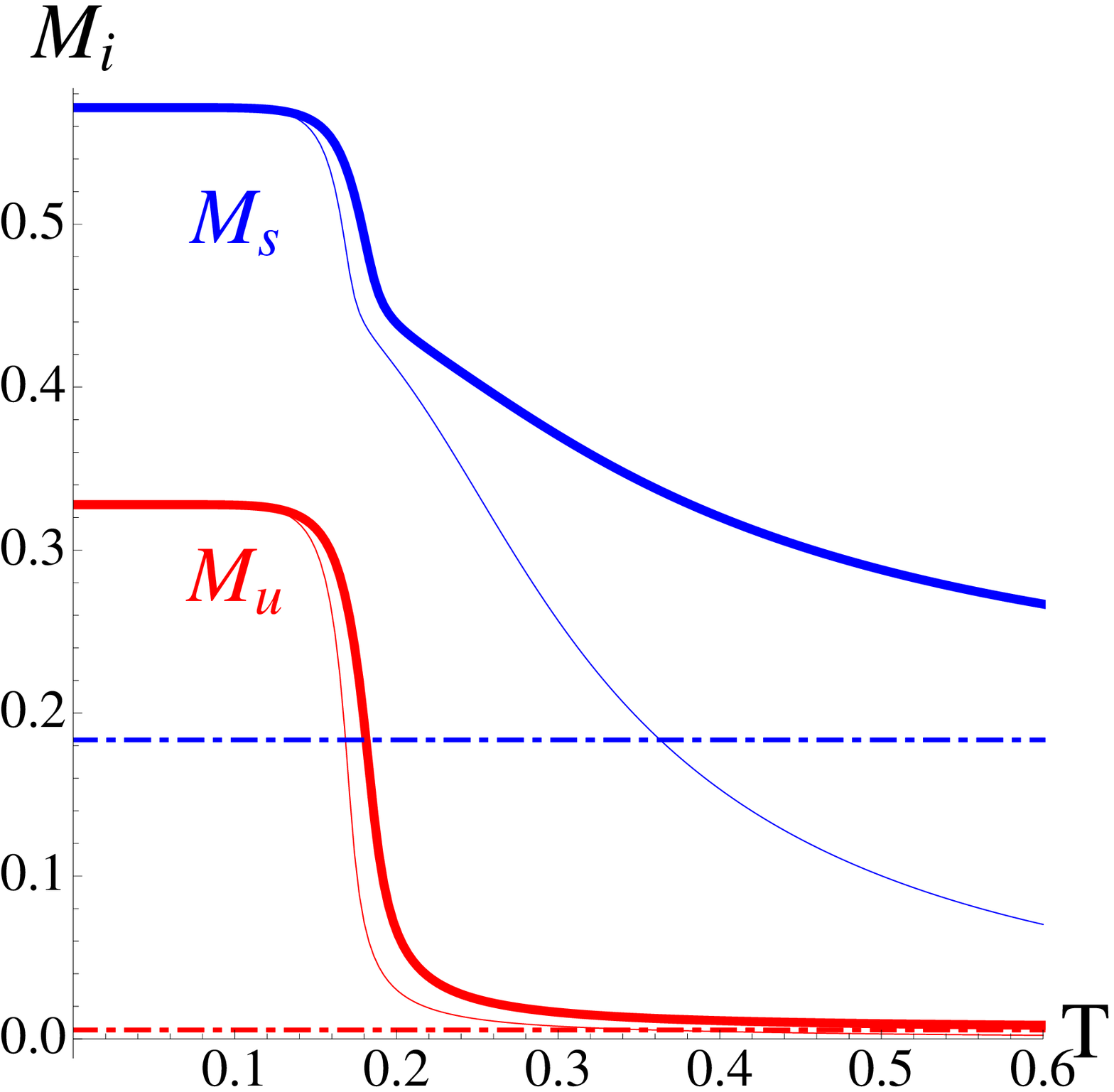}}
\caption{
The same as in the previous two Figures  (same notation for the curves as well) but using the potential $\mathcal{U}^{IV}$ (see Table \ref{PolyakovPot}, $T_0=0.19~\mathrm{GeV}$ and with $\kappa=0.06$): thick lines using set (III) from Table \ref{ParamSets}  and thin lines using set (IV) from Table \ref{ParamSets}.}
\end{center}
\end{figure}

\vspace{0.5cm}

\section {Concluding remarks}

We conclude highlighting the main results:

1- We have presented a way to obtain the thermodynamic potential by integrating the model gap equations and making use of an integration constant to ensure the correct low $T,\mu$ behaviour of the quark number degrees of freedom. At the same time the Stefan-Boltzmann limit is correctly described.
The quark one loop amplitudes obtained in this scheme are calculated using the Pauli-Villars regularization acting both on the divergent integrals related to the Dirac vacuum and on the finite integrals over Fermi distributions. A {\it simultaneous} description of thermodynamic observables related with the NJL models and extensions thereof is accomplished. We addressed three observables: the number of degrees of freedom, the quark mass solutions (and implicitly of the condensates) and the traced Polyakov loop solutions of the respective gap equations. In particular the large $T$ asymptotic behavior of the number of degrees of freedom related with multiquark versions of the NJL model, NJLH and NJLH8, as well as their Polyakov extended versions PNJLH and PNJLH8, is correctly described. This is one of our main results. As is well known, in the case of the conventional sharp 3D regularization the Stefan-Boltzman limit in NJL and PNJL models is obtained only if the matter parts are not regulated. 
This choice leads however to an incorrect asymptotics for the quark mass and traced Polyakov loop gap solutions as indicated in the next point.  

2- We obtain that the constituent quark masses and the Polyakov classical fields $\phi,\overline{\phi}$ calculated with  the PV as well as the 3D sharp cutoff regularizations lead to similar consistent asymptotic results, when the Fermi distributions are regularized. The quark masses approach the current quark mass values at large $T$. The corresponding condensates vanish. As opposed to this a removal of the regulator from the matter parts would result in vanishing quark masses in the large $T$ limit, which in turn are associated with unphysical positive condensates. 
Regarding the $\phi,\overline{\phi}$ fields, we show that the Polyakov potential $\mathcal U$ drives the large $T$ asymptotic behavior of the $\phi^*,\overline{\phi}^*$ gap equation solutions when the matter integrals are regulated. Consequently the "overshooting" of these variables at large $T$ above the values of the minima of $\mathcal U$ in the examples shown is consequence of inadequately removing the cutoff from the finite matter integrals. 
  
3- The fact that in point 1 only the PV regularization of vacuum and matter integrals leads to a consistent description of the asymptotic behavior, while in point 2 also the 3D regularization applied to vacuum and matter parts works, has a very simple reason. The constituent quark masses and $\phi,\overline{\phi}$ can only detect effects related with the gap equations, while the number of degrees of freedom are highly sensitive to the $T,\mu$ dependence of the $M_f$ and $\phi,\overline{\phi}$ independent integration constant $C(T,\mu)$ (therefore irrelevant for the gap equations) which is associated with the proposed procedure to derive the thermodynamic potential through integration of the gap equations.  

4- Finally, it is possible to extend the method presented here to the 3D cutoff procedure (see Appendix C)  and in this case both the results of PV and 3D regularizations will be compatible with each other with the same degree of thermodynamic consistency for the observables discussed. 
 

\section {Appendix}

{\bf A.} Analytical expressions for the large $T$ limit of $J^{med}_{-1}(M_f^2,T,\mu,\phi,\overline{\phi} )$ integral in specific cases:

{\bf a)} $\mu=0,\overline{\phi}=\phi=1$; 

The pertinent integral for this case is eq. (\ref{int4})

\begin{equation} 
\label{Ap1}
  I =   \int^\infty_0\ud |\vec{p}_E||\vec{p}_E|^4\frac{n_{qM}}{E_p(M)}
     =T^4\int^\infty_0\!\ud x\frac{(x^2+2x\frac{M}{T})^{\frac{3}{2}}}{1+
     e^{x+\frac{M}{T}}}
\end{equation}

With the variable transformation $y^2=x^2 +2x\frac{M}{T}$ it can be rewritten as

\begin{equation} 
\label{Ap2}     
  I(M^2,T)=   T^4\int^\infty_0\!\ud y\frac{y^4}{(\sqrt{y^2+a^2})(1+
     e^{\sqrt{y^2+a^2}})}
\end{equation} 
with $a^2=\frac{M^2}{T^2}$. The large $T$ limit can be studied by Taylor expanding around $a^2=0$, in complete analogy to  \cite{Jackiw:1974}. The leading and subleading orders are easy to obtain, using
\begin{equation}
I(M^2,T)|_{a^2=0}=\frac{7 \pi^4}{120}T^4 
\end{equation}
and  
\begin{eqnarray}
\frac{d I(M^2,T)}{d a^2}|_{a^2=0}&=T^4 &\int^\infty_0\ud y y^4\frac{d}{d y^2}\frac{1}{(\sqrt{y^2+a^2})(1+
     e^{\sqrt{y^2+a^2}})}|_{a^2=0}\nonumber \\
&=&-\frac{3 T^4}{2}\int^\infty_0\ud y \frac{y^2}{(\sqrt{y^2+a^2})(1+
     e^{\sqrt{y^2+a^2}})}|_{a^2=0}=-T^4\frac{\pi^2}{8},
\end{eqnarray}
where a partial integration was done to obtain the last integral.
The remaining terms 
\begin{eqnarray}
\frac{d^2 I(M^2,T)}{(d a^2)^2}&=&-\frac{3 T^4}{2}\int^\infty_0\ud y y^2\frac{d}{d y^2}\frac{1}{(\sqrt{y^2+a^2})(1+
     e^{\sqrt{y^2+a^2}})}\nonumber \\
&=&\frac{3}{4}T^4 \int^\infty_0\ud y \frac{1}{(\sqrt{y^2+a^2})(1+
     e^{\sqrt{y^2+a^2}})}
\end{eqnarray}
are identical to the non analytic contribution evaluated in \cite{Jackiw:1974}. 
We therefore just quote the final result for the high $T$ limit
\begin{equation}
I(M^2,T)= \frac{7 \pi^4}{120} T^4 -\frac{\pi^2}{8} M^2 T^2 -\frac{3}{32}M^4ln[M^2T^{-2}]-\frac{3}{32}M^4 c +{\cal O}(M^6T^{-2})
\end{equation}
with $c=2 \gamma -\frac{3}{2} -2 ln \pi$.
\vspace{0.5cm}
At $M=0$, $I(0,T)=\frac{7 \pi^4}{120} T^4$.
Applying the Pauli-Villars operator to these type of integrals appearing in(\ref{J-1med}), we obtain that the regularized integral
\begin{eqnarray}
\label{equ} 
&&J^{med}_{-1}(M_f^2,T\rightarrow \infty,\mu=0,\phi=1,\overline{\phi}=1)\nonumber \\
&=&-\frac{16}{3}[I(M_f^2,T)-I(M_f^2+\Lambda^2,T)+\Lambda^2\frac{d}{d\Lambda^2}I(M_f^2+\Lambda^2,T)]\nonumber \\
&+&\frac{16}{3}[I(0,T)-I(\Lambda^2,T)+\Lambda^2\frac{d}{d\Lambda^2}I(\Lambda^2,T)]\nonumber \\
&=&\frac{1}{2} (M_f^4 ln(\frac{M_f^2}{M_f^2+\Lambda^2})+\Lambda^4 ln(\frac{M_f^2+\Lambda^2}{\Lambda^2}) +\Lambda^2 M_f^2)\nonumber \\
&=&\frac{1}{2}(M_f^2 J_0(M_f^2)+\Lambda^4 ln(1+\frac{M_f^2}{\Lambda^2})).
\end{eqnarray}
reduces to the negative of the vacuum expression in the $T\rightarrow \infty$ limit, eq. (\ref{J_1}).  The large $T$ asymptotics of the thermodynamic potential is therefore dictated by the $M$ independent $C(T,\mu)$ term.

As mentioned in section II one sees that in the leading order of the high $T$ expansion terms proportional to $I(0,T)$ and $I(M,T)$ are equal to each other and to the $C(T,\mu)$ term. The associated PV regulating masses cancel in the leading order as well. 
\vspace{0.5cm}

{\bf b)} $\mu=0, \phi\ne 1,\overline{\phi}\ne 1$

From (\ref{occpol}) one obtains after a variable substitution 
\begin{eqnarray}
&&I[{\phi},\overline{\phi}]=\int^\infty_0\ud |\vec{p}_E||\vec{p}_E|^4 \frac{1}{{E_p(M)}}(\tilde{n}_{qM}\left(E_p,\mu=0,T,\phi,\overline{\phi}\right)+\tilde{n}_{\overline{q}M}\left(E_p,\mu=0,T,\phi,\overline{\phi}\right))\nonumber \\
&=&T^4\int^\infty_0\ud y y^3 \frac{1+ e^{2 \sqrt{(y^2+a^2)}} \phi +2  e^{ \sqrt{(y^2+a^2)}}  \overline{\phi}}{1+e^{3 \sqrt{(y^2+a^2)}}+3 e^{2 \sqrt{(y^2+a^2)}}\phi +3 e^{ \sqrt{(y^2+a^2)}} \overline{\phi}} + \phi \leftrightarrow \overline{\phi}
\end{eqnarray}
which is part of the $J^{med}_{-1}(M_f^2,T,\mu,\phi,\overline{\phi} )$ integral.
As before, the first two orders in an expansion around $a^2=0$ lead to regular integrals. For the present discussion it is important to realize that the leading term, proportional to $T^4$ is a $\phi,\overline{\phi}$ dependent function, for example along $\phi=\overline{\phi}$ one obtains the closed form expression
\begin{equation}
I[\phi,\phi]|_{a^2=0}=T^4 [\frac{7 \pi^4}{360}-2 L_{i4}(\frac{2}{1-3 \phi +\sqrt{(9 \phi^2 -6 \phi-3)}}) -2 L_{i4}(\frac{2}{1-3 \phi -\sqrt{(9 \phi^2 -6 \phi-3)}})], 
\end{equation}
where $L_{i4}(x)$ designates the polylogarythm function of 4th order. Therefore the asymptotics of the unregularized integral $I[{\phi},\overline{\phi}]$, contributes to the number of degrees of freedom associated with the Polyakov loop. 
This is to be contrasted to the regularized $J^{med}_{-1}(M_f^2,T,\mu,\phi,\overline{\phi} )$, where in analogy to the example a) the $T^4$ dependence cancels out, being restored through the ${\phi},\overline{\phi}$ independent $C(T,\mu)$ term. 

The leading order in $a^2$ of $J^{med}_{-1}(M_f^2,T,\mu,\phi,\overline{\phi} )$ is the only term that contributes in the limit of vanishing current quark masses after the chiral transition has taken place, see Fig. 7 below in part B of the Appendix, and related discussion. In this case the results for the gap equations for $\phi,\overline{\phi} $ are known analytically in the $T\rightarrow\infty$ limit and serve to illustrate some features of the asymptotics of the more general cases considered in section V. 
Supposing once again that the medium part is not regularized, one obtains in the chiral limit and for vanishing chemical potential, for $T>T_c$ that 

\begin{equation}
\frac{dI[{\phi},\overline{\phi}]}{d \phi}|_{\overline{\phi}=\phi}= 2 T^4 \frac{ [ L_{i3}(\frac{2}{1-3 \phi -\sqrt{(9 \phi^2 -6 \phi-3)}}) - L_{i3}(\frac{2}{1-3 \phi +\sqrt{(9 \phi^2 -6 \phi-3)}})]}{\sqrt{(9 \phi^2 -6 \phi-3)}},
\end{equation}
with $L_{i3}(x)$ the polylogarythm function of 3rd order, and similarly for the derivative with respect to 
$\overline{\phi}$  due to the symmetry under $\phi \leftrightarrow \overline{\phi}$ of the thermodynamic potential in absence of $\mu$. 
Since the hight $T$ dependence is the same as for the derivatives of the Polyakov potential, one obtains that the asymptotic values of $\phi$ are also steered by the medium part of the quark loop integral in presence of the Polyakov loop. In particular as $\phi\rightarrow 1$ the contribution to the gap equation for the $\phi$ variable stemming from the quark loop integral is  $\frac{-16}{3}\frac{1}{T^4}\frac{dI[{\phi},\overline{\phi}]}{d \phi}|_{\overline{\phi}=\phi=1}= \frac{-8\pi^2}{3} $. Thus it is evident that $\phi=1$ cannot be the asymptotic solution of the gap equation for $\phi$ for polynomial Polyakov potentials that display a minimum at $\phi=1$ in the large $T$ limit. 
\vspace{0.5cm}

{\bf B.} The regularization dependence of the quark mass gap equation.
\vspace{0.5cm}

We illustrate in Fig. 6 for the $SU(3)$ flavor limit the behavior of the  regularization dependence of the solutions of the quark mass gap equation  by plotting the quark one-loop integral $J_0(M^2,T,\mu)$ of eq. (\ref{gap-t}) for different temperatures (solid lines) and the $T$ independent quantity
\begin{equation}
\label{hoM}
-\frac{2\pi^2 h}{N_c M}=-\frac{2\pi^2}{N_c}[\frac{m}{h}-(G+\frac{\kappa}{16} h +\frac{3}{4} (g_1+\frac{2}{3}g_2) h^2]^{-1}, 
\end{equation}
represented by the dashed line, as function of the condensate $h$. We use  the SPA eqs. (\ref{StaEq}) to express $M$ as function of $h$, since there is a one to one correspondence between these quantities for the model parameters that ensure the existence of a stable effective potential. The intersection of the dashed line and solid lines indicates the solution in the dynamically broken phase of chiral symmetry at a certain temperature (here we considered vanishing chemical potential and set $\phi=\overline{\phi}=1$). Physical solutions have negative or zero-valued condensates $h$. In absence of an intersection only the trivial solution $M=0$ survives (not indicated in the figure, since we are plotting only the non-trivial solutions, as we take the ratio $\frac{h}{M}$.
In the chiral limit (left panel) this quantity is bounded, has a continuous behavior and in the limit of vanishing condensate we have $\lim_{h\rightarrow 0}\frac{h}{M}=-\frac{1}{G}\ne 0$. The inclusion of a finite quark mass however induces a pole (as can be seen in the middle and right panels) which for a physically relevant choice of model parameters lies in the positive $h$ side (note that this quantity, expressed as a function of $h$ only depends on the choice of the current mass and on the coupling strengths) and the ratio goes to zero with vanishing quark condensate. The solid lines correspond to the evaluation $J_0$ for different temperatures. The rightmost panel corresponds to the case where the medium part is not regularized, whereas in the other two all contributions are PV regularized. In all of them the increase in temperature corresponds to an increase in the negative contribution coming from the medium part. From top to bottom the lines correspond to: $T=0, 0.1, 0.2, 0.3, 0.4~\mathrm{GeV}$.
 
One can see that in the chiral limit the intersection ceases to exist above a certain critical temperature leaving only the trivial solution ($M=0$). With the inclusion of a finite current mass there is always an intersection because of the divergence near the pole. In the fully regularized version the integral tends to zero with increasing temperature and as a result the intersection tends to the solution $h=0$ ($M=m$). When the medium contribution is unregularized the integral diverges with the temperature and as a result the intersection tends to the pole ($M=0$).

\begin{figure}[htp]
\begin{center}
\label{intersect}
\subfigure[]{\label{graf4D}\includegraphics[width= 0.3 \columnwidth]{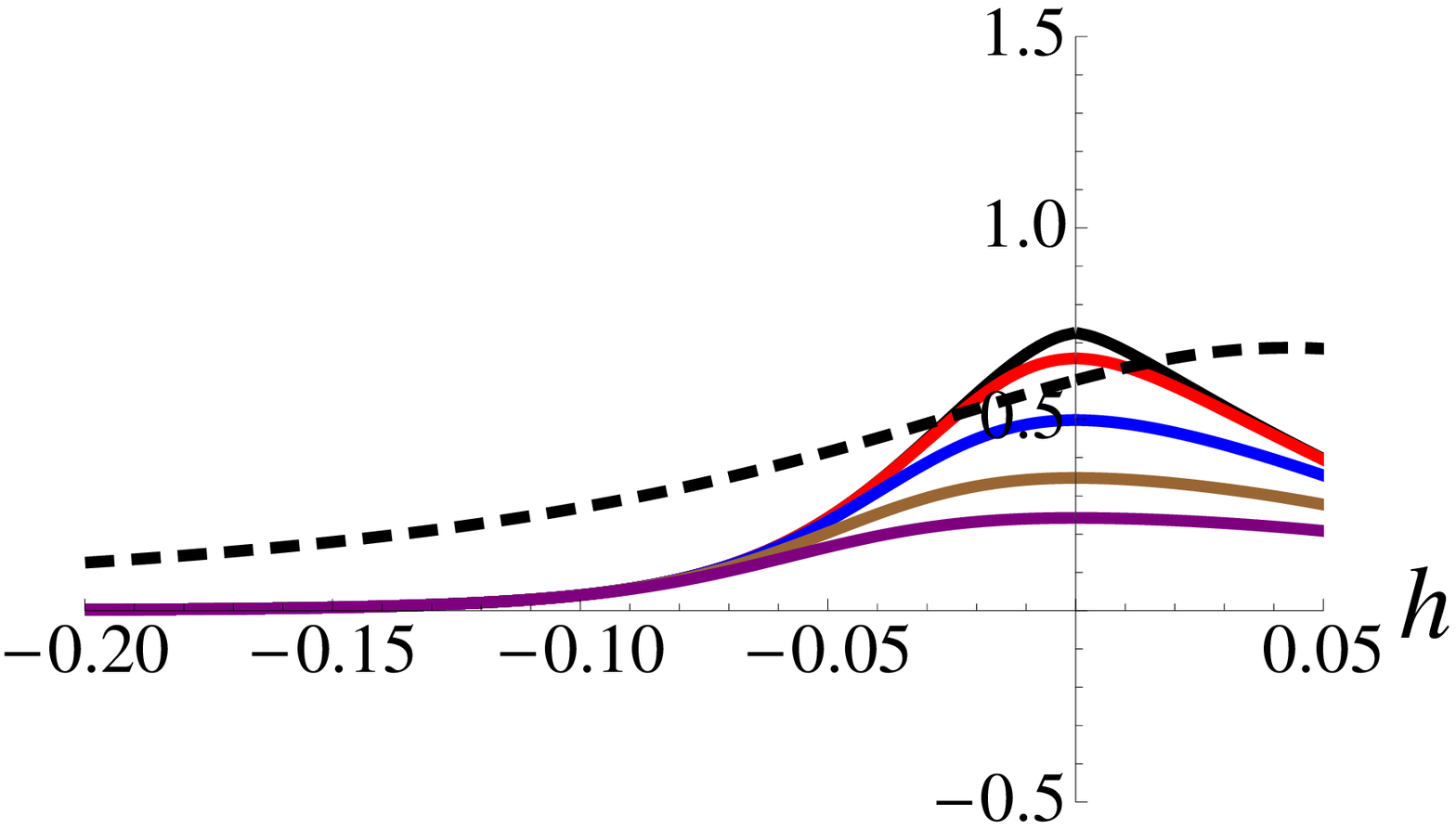}}
\subfigure[]{\label{graf4Dm}\includegraphics[width= 0.3 \columnwidth]{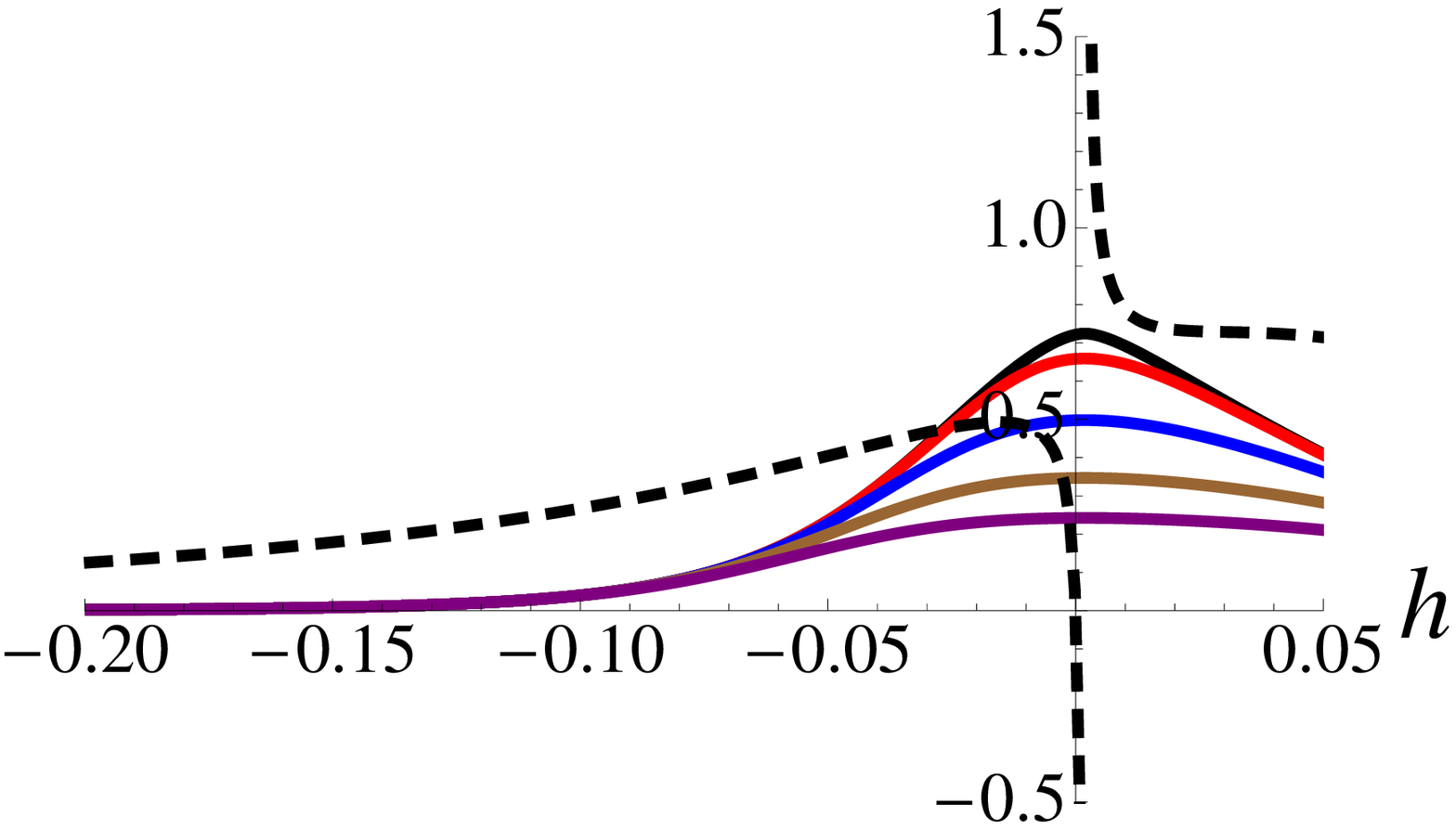}}
\subfigure[]{\label{graf4DmSemCutoff}\includegraphics[width= 0.3 \columnwidth]{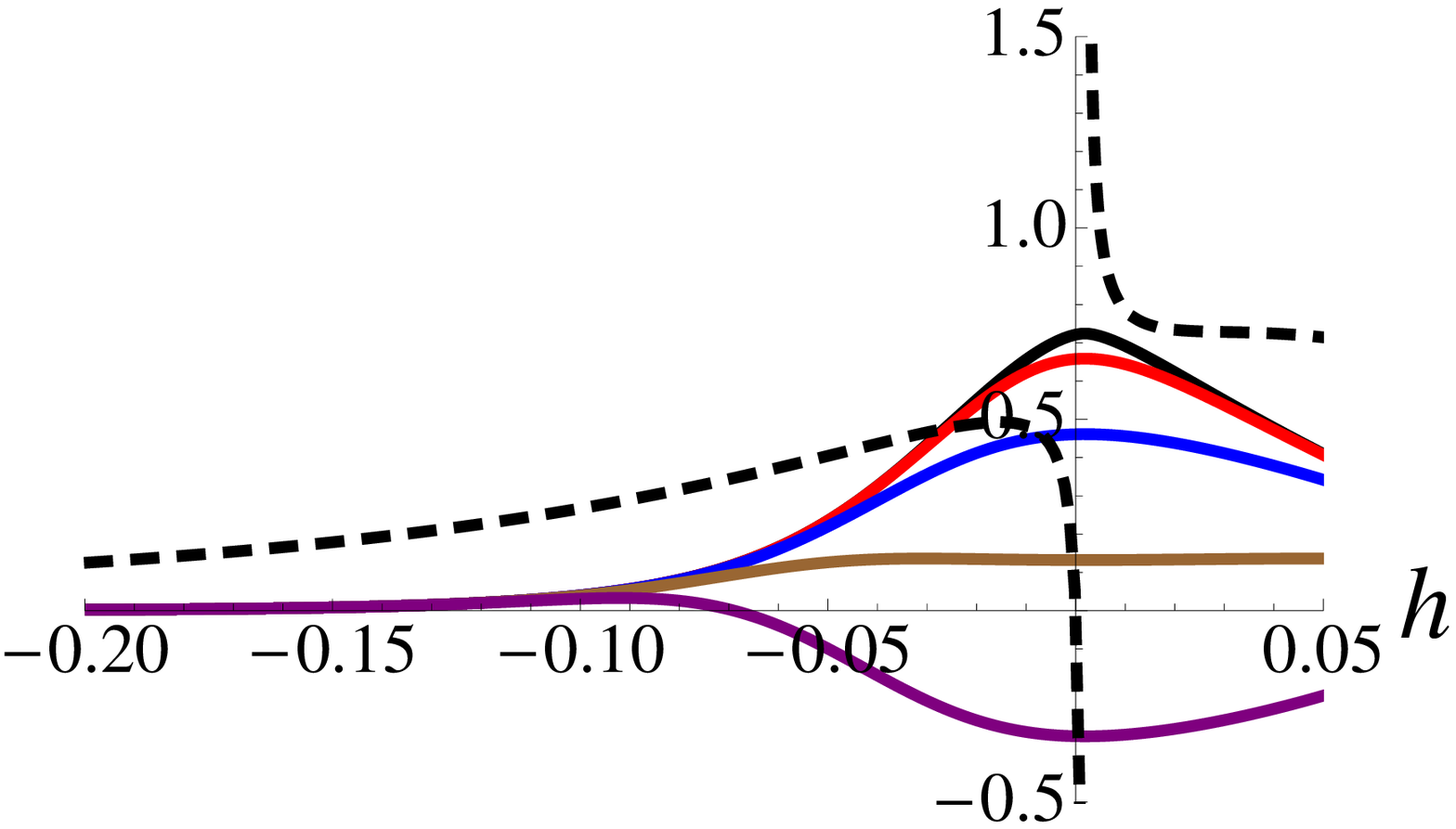}}
\caption{
The intersection of the quark loop integral $J_0(M^2,T,\mu=0)$ represented at different temperatures  by the solid lines (top for $T=0$, lower ones for increasing $T=0.1, 0.2, 0.3, 0.4~\mathrm{GeV}$) with the $T$ independent ratio $\frac{h}{M}$ given in (\ref{hoM}) (dashed line) corresponds to a non-trivial solution of the gap equation (the vertical axis is in $\mathrm{GeV}^2$ units). All curves are plotted as functions of the condensate $h$. Left panel: current quark mass $m$=0 and overall PV regularization. Middle: $m\ne 0$ and overall PV regularization. Right: $m\ne 0$ removing the PV regulator from the matter parts. All cases are for the $SU(3)$ flavor limit with the parameters of set III from Table \ref{ParamSets} except for the current quark masses which were chosen to be $m=0~\mathrm{MeV}$ and $m=20~\mathrm{MeV}$. See discussion in the text. } 
\end{center}
\end{figure}

In Fig. 7 we show that in presence of the Polyakov loop and for non-vanishing chemical potential the pattern of the solutions for the $SU(3)$ chiral limit with overall PV regularization (left panel) is still similar to the case presented on the left of Fig. 6. Removing the cutoff from the matter part (right panel) changes the behavior of the solid curves only at somewhat higher temperatures. In this case, since the intersection ceases to exist at low temperatures, i.e. a first order transition takes place, the effect of removing the regulator is not noticeable in the solutions. Therefore in our study of the regularization effects on the gap equation solutions for the Polyakov loop at high temperature in the chiral limit we can start in the phase where the chiral transition has taken place for the quark mass.

\begin{figure}[htp]
\begin{center}
\label{intersect2}
\subfigure[]{\label{graf4Dmuphiphibar}\includegraphics[width= 0.3\columnwidth]{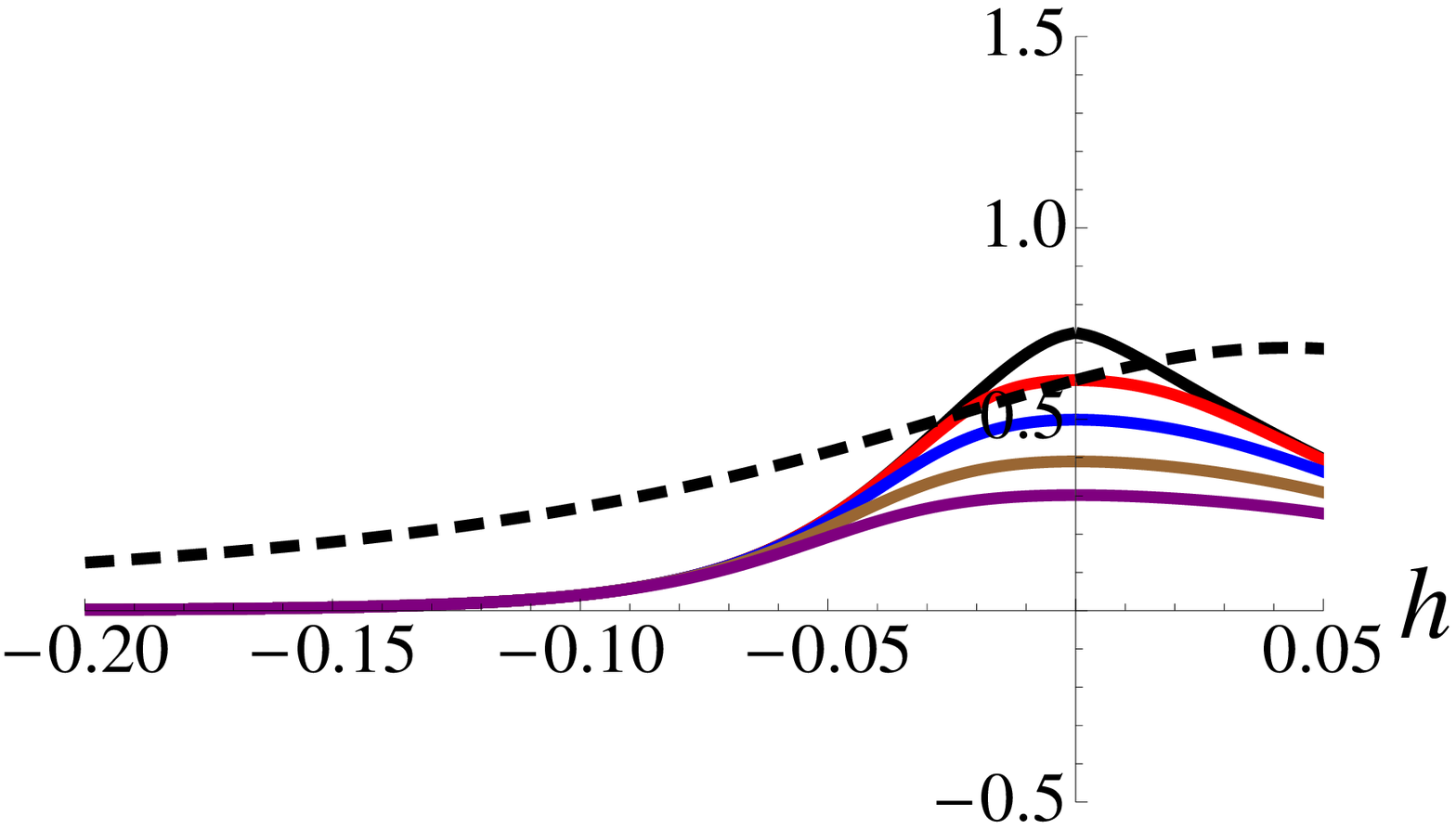}}
\subfigure[]{\label{graf4DmuphiphibarSemCutoff}\includegraphics[width= 0.3 \columnwidth]{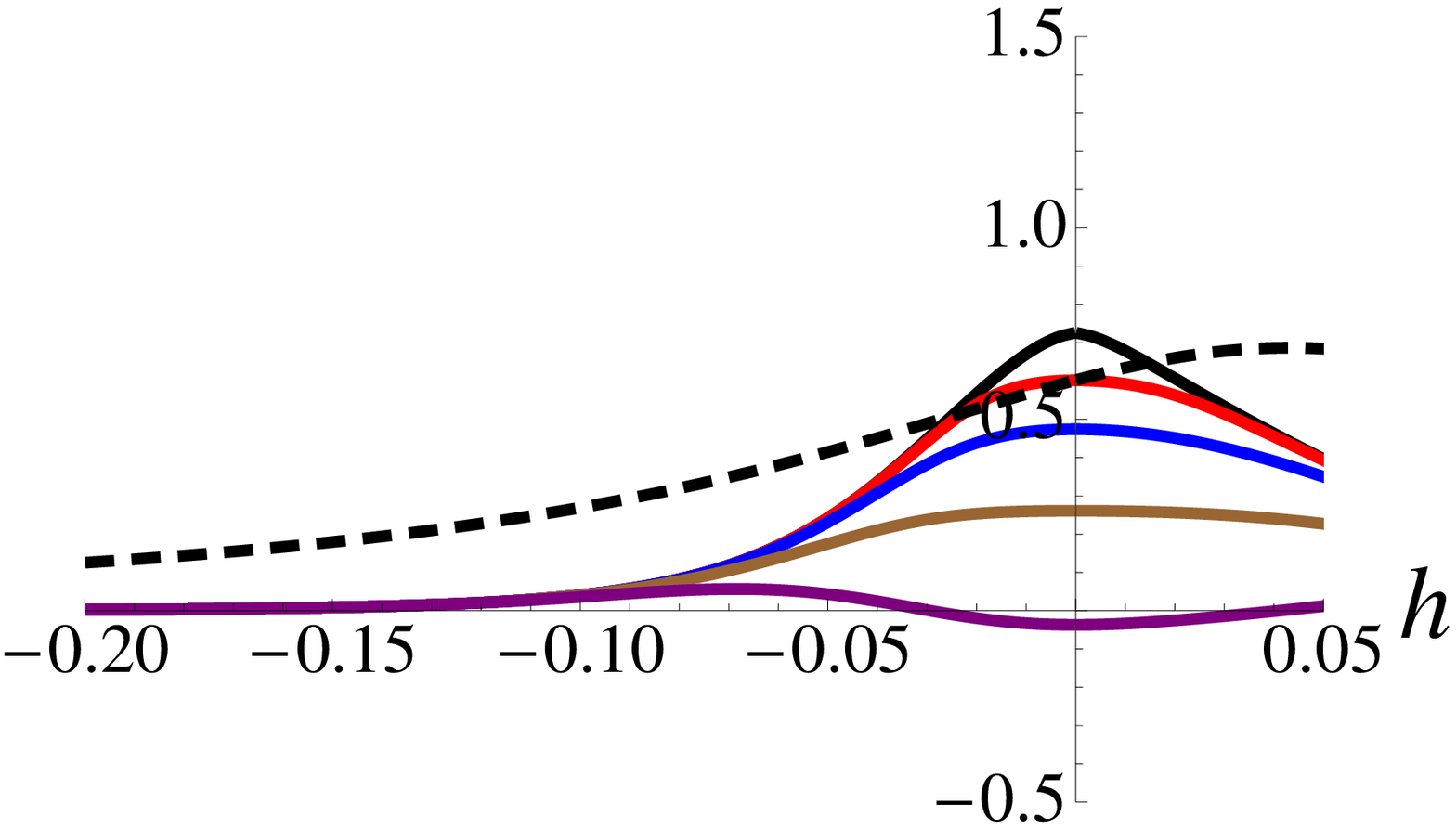}}
\caption{
The quark loop integral $J_0(M^2,T,\mu,\phi,\overline{\phi})$ of the PNJL in the chiral limit and the ratio $\frac{h}{M}$ given in (\ref{hoM}), using the same notation as in the previous Figure and with the same parameters. Here we consider however finite value for the chemical potential and the Polyakov loop ($\mu=0.2,~\phi=0.4,~\overline{\phi}=0.6$). Left: overall PV regularization. Right: the PV regulator is removed from the matter parts. The intersection of solid lines with the dashed line yield a solution to the gap equation for the quark mass. See discussion in the text for further details. } 
\end{center}
\end{figure}
\vspace{1.5 cm}

{\bf C.} The 3D sharp-cutoff case extended.
\vspace{.5 cm}

The pertinent Feynman amplitudes $J_0(M^2,T,\mu )$, eq. (\ref{J0t}), and $J_{-1}(M^2,T,\mu )$, eqs. (\ref{J-1}),(\ref{J_1}),(\ref{J-1med}) are conveniently expressed in terms of the PV regularization kernel (\ref{PV}), which can be readily exchanged by other regularization kernels. We consider now the case of the 3D sharp cutoff regulator ${\hat \rho}_{\Lambda_3} = 1-\Theta(|\vec p_E|-\Lambda_3)$  
where the first term leaves the integrand unmodified in similarity with the PV regulator. The PV subtractions are then to be replaced by the second term of ${\hat \rho}_{\Lambda_3}$. 
This leads to the following results for the integrals considered:

(i) it is easy to verify that the following vacuum integral which is part of $J_{-1}$  without zero mass subtraction is finite, as opposed to the PV case, but non-zero at $M=0$
\begin{eqnarray}
&&\int_0^{\infty}\ud |\vec{p}_E| |\vec{p}_E|^2 \hat{\rho}_{\Lambda_3} E_p(M) \nonumber \\ 
&=&
\frac{1}{8} (\Lambda_3 \sqrt{(\Lambda_3^2+M^2)} (2 \Lambda_3^2 + M^2) + M^4 ln [M/(\Lambda_3+\sqrt{\Lambda_3^2+M^2})]). 
\end{eqnarray}

As such a zero mass subtraction is still required for the effective potential to start off at zero for $M=0$. 

(ii) Regarding the medium integral (\ref{J-1med}) one has, keeping track of the non vanishing surface term in this case,
\begin{eqnarray}
\label{J-1ap}
     J_{-1}^{\mathrm{med_3}}(M^2,T,\mu )&=&
\frac{8 T \Lambda_3^3}{3}ln\left[\frac{(1+e^{(-\frac{\Lambda_3 -\mu}{T})})(1+e^{(-\frac{\Lambda_3 +\mu}{T})}))}{(1+e^{(-\frac{\sqrt{(\Lambda_3^2+M^2)} -\mu}{T})})(1+e^{(-\frac{\sqrt{(\Lambda_3^2+M^2)} +\mu}{T})})}\right]\nonumber \\
&&-\frac{8}{3}\int^\infty_0\ud 
     |\vec{p}_E||\vec{p}_E|^4\hat{\rho}_{\Lambda_3}\left(
     \frac{n_{qM}+n_{\overline{q}M}}{E_p(M)}-
     \frac{n_{q0}+n_{\overline{q}0}}{E_p(0)}\right).
\end{eqnarray}
The surface term vanishes in the limit $T\rightarrow 0$, but at  $T\rightarrow \infty$ it behaves (for $\mu=0$) as $-\Lambda_3^4 +\Lambda_3^3 \sqrt{\Lambda_3^2+M^2} -\frac{\Lambda_3^3 M^2}{T}+ {\cal O}(1/T^2)$ and a small dependence on the model parameters of the order of $M^2/\Lambda_3^2\rightarrow m^2/\Lambda_3^2$ (where $m$ is the current quark mass, to which the constituent quark mass approaches asymptotically) remains at leading order of the expansion, as opposed to the case of PV regularization, where the boundary term is absent. However this will not affect the high $T$ asymptotics, which will be lead by the $C(T,\mu)\sim T^4$ term, as in the PV case, see below.
The remaining integral behaves qualitatively like in the PV case, with the same leading order behavior in the low and high $T$ expansions. To see this, we use again representation (\ref{Ap1}) to obtain (at $\mu=0$)
\begin{equation} 
\label{Ap1a}
     \int^\infty_0\ud |\vec{p}_E||\vec{p}_E|^4 \hat{\rho}_{\Lambda_3} \frac{n_{qM}}{E_p(M)}
     =T^4\left(\int^\infty_0\!\ud x\frac{(x^2+2x\frac{M}{T})^{\frac{3}{2}}}{1+
     e^{x+\frac{M}{T}}}-\int^\infty_{(M+\sqrt{M^2+\Lambda_3^2})/T}\!\ud x\frac{(x^2+2x\frac{M}{T})^{\frac{3}{2}}}{1+
     e^{x+\frac{M}{T}}}\right)
\end{equation}
The regulator dependent piece, the second integral, differs from the corresponding expression for the PV subtractions, but also vanishes exponentially with $T$ as $T\rightarrow 0$ for non zero values of $M$ and $\Lambda$ as in the PV case. Therefore, as in the PV case the low $T$ behavior stems again from the same unphysical zero mass occupation states (the regulator independent part of the second integral of (\ref{J-1ap})); the $C(T,\mu)$ term, eq. (\ref{FTP}), is used analogously to correct it. Therefore the low  and high $T$ behavior will be the same as in the PV case. 

In Fig. 8 are shown the degrees of freedom for the PNJL model calculated with the advocated way of obtaining the thermodynamic potential, using the PV and the 3D sharp cutoff regulators. One sees that the low and high $T$ regimes coincide. In the crossover region the number of degrees of freedom is higher for the 3D cutoff as for the PV regulator. This is in contrast with what one observes comparing the PV regularization with the conventional 3D treatment and the regulator removed from the matter parts, where the behavior is opposite.   

\begin{figure}[thp]
\begin{center}
\label{3DCorr}
\includegraphics[width= 0.3 \columnwidth]{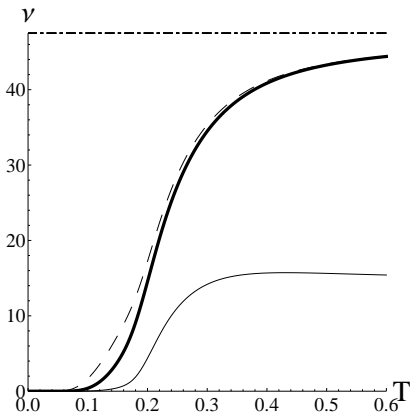}
\caption{Temperature dependence of the effective degrees of freedom using when the regulator is kept in all contributions: thick line corresponds to PV regularization, thin line to the usual 3D cutoff procedure and dashed line to the above discussed extension of 3D regularization. The choice of parameters is the same as described in Fig. 1.}
\end{center}
\end{figure}
\vspace{1.5cm}
    
\centerline{\bf ACKNOWLEDGMENTS}
\vspace{0.5cm}

This  work  has  been  supported  in  part  by  grants  of Funda\cc \~ao para a Ci\^encia e Tecnologia,  FEDER,  OE,
SFRH/BPD/63070/2009, CERN/FP/116334/2010 and Centro de F\'isica Computacional, unit 405. We acknowledge  the support of the  European Community-Research  Infrastructure  Integrating  Activity  Study  of Strongly  Interacting  Matter  (acronym  HadronPhysics2, Grant   Agreement   No.   227431)   under   the   Seventh Framework Programme of the EU.

\end{document}